\documentclass[12pt]{article}

 \usepackage{amssymb}   %%% note: needed for \mathbb{}, \dal=\square
 \usepackage{graphicx}
 \usepackage{cite}

\begin{document}

%%%%%%%%%%%%%%%%%%%%%%%%%%%%%%%%%%%%%%%%%%%%%%%%%%%%%%%%%%%%%%%%%%%%%%%%%%%%%%%

\begin{center}
{\LARGE\bf
Eigenvalue based taste breaking of staggered,\\[1mm]
Karsten-Wilczek and Bori\c{c}i-Creutz fermions\\[3mm]
with stout smearing in the Schwinger model}
\end{center}

\vspace{20pt}

\begin{center}
{\large\bf Maximilian Ammer$\,^{a}$}
\,\,and\,\,
{\large\bf Stephan D\"urr$\,^{a,b}$}
\\[10pt]
${}^a${\sl Department of Physics, University of Wuppertal, 42119 Wuppertal, Germany}\\
${}^b${\sl J\"ulich Supercomputing Centre, Forschungszentrum J\"ulich, 52425 J\"ulich, Germany}
\end{center}

\vspace{10pt}

\begin{abstract}
In two spacetime dimensions staggered fermions are minimally doubled, like Karsten-Wilczek and Bori\c{c}i-Creutz fermions.
A continuum eigenvalue is thus represented by a pair of near-degenerate eigenvalues, with the splitting $\delta$ quantifying the cut-off induced taste symmetry breaking.
We use the quenched Schwinger model to determine the low-lying fermionic eigenvalues (with 0, 1 or 3 steps of stout smearing), and analyze them in view of the global topological charge $q\in\mathbb{Z}$ of the gauge background.
For taste splittings pertinent to would-be zero modes, we find asymptotic Symanzik scaling of the form $\delta_\mathrm{wzm} \propto a^2$ with link smearing, and $\delta_\mathrm{wzm} \propto a$ without, for each action.
For taste splittings pertinent to non-topological modes, staggered splittings scale as $\delta_\mathrm{ntm} \propto a^p$ (where $p\simeq2$ with smearing and $p=1$ without),
while Karsten-Wilczek and Bori\c{c}i-Creutz fermions scale as $\delta_\mathrm{ntm} \propto a$ (regardless of the smearing level).
Large logarithmic corrections are seen with smearing.
\end{abstract}

\vspace{20pt}

\newcommand{\pad}{\partial}
\newcommand{\hqu}{\hbar}
\newcommand{\til}{\tilde}
\newcommand{\pri}{^\prime}
\renewcommand{\dag}{^\dagger}
\newcommand{\<}{\langle}
\renewcommand{\>}{\rangle}
\newcommand{\gaf}{\gamma_5}
\newcommand{\nab}{\nabla}
\newcommand{\lap}{\triangle}
\newcommand{\dal}{\square}
\newcommand{\trc}{\mathrm{tr}}
\newcommand{\Trc}{\mathrm{Tr}}
\newcommand{\Mpi}{M_\pi}
\newcommand{\Fpi}{F_\pi}
\newcommand{\Mka}{M_K}
\newcommand{\Fka}{F_K}
\newcommand{\Met}{M_\et}
\newcommand{\Fet}{F_\et}
\newcommand{\Mss}{M_{\bar{s}s}}
\newcommand{\Fss}{F_{\bar{s}s}}
\newcommand{\Mcc}{M_{\bar{c}c}}
\newcommand{\Fcc}{F_{\bar{c}c}}

\newcommand{\al}{\alpha}
\newcommand{\be}{\beta}
\newcommand{\ga}{\gamma}
\newcommand{\de}{\delta}
\newcommand{\ep}{\epsilon}
\newcommand{\ve}{\varepsilon}
\newcommand{\ze}{\zeta}
\newcommand{\et}{\eta}
\renewcommand{\th}{\theta}
\newcommand{\vt}{\vartheta}
\newcommand{\io}{\iota}
\newcommand{\ka}{\kappa}
\newcommand{\la}{\lambda}
\newcommand{\rh}{\rho}
\newcommand{\vr}{\varrho}
\newcommand{\si}{\sigma}
\newcommand{\ta}{\tau}
\newcommand{\ph}{\phi}
\newcommand{\vp}{\varphi}
\newcommand{\ch}{\chi}
\newcommand{\ps}{\psi}
\newcommand{\om}{\omega}

\newcommand{\qhat}{\hat{q}}
\newcommand{\khat}{\hat{k}}

\newcommand{\bdm}{\begin{displaymath}}
\newcommand{\edm}{\end{displaymath}}
\newcommand{\bea}{\begin{eqnarray}}
\newcommand{\eea}{\end{eqnarray}}
\newcommand{\beq}{\begin{equation}}
\newcommand{\eeq}{\end{equation}}

\newcommand{\mr}{\mathrm}
\newcommand{\mb}{\mathbf}
\newcommand{\ri}{\mr{i}}
\newcommand{\Nf}{N_{\!f}}%{{N_{\!f}}}
\newcommand{\Nc}{N_{ c }}%{{N_{ c }}}
\newcommand{\Nt}{N_{ t }}%{{N_{ t }}}
\newcommand{\Nv}{N_{ v }}%{N_\mr{vec}}
\newcommand{\Nthr}{N_\mr{thr}}
\newcommand{\DW}{D_\mr{W}}
\newcommand{\DB}{D_\mr{B}}
\newcommand{\DN}{D_\mr{N}}
\newcommand{\DS}{D_\mr{st}}%{D_\mr{S}}
\newcommand{\DA}{D_\mr{A}}
\newcommand{\DKW}{D_\mr{KW}}
\newcommand{\DBC}{D_\mr{BC}}
\newcommand{\HW}{H_\mr{W}}
\newcommand{\HB}{H_\mr{B}}
\newcommand{\HN}{H_\mr{N}}
\newcommand{\HS}{H_\mr{S}}
\newcommand{\HA}{H_\mr{A}}
\newcommand{\MeV}{\,\mr{MeV}}
\newcommand{\GeV}{\,\mr{GeV}}
\newcommand{\fm}{\,\mr{fm}}
\newcommand{\MSbar}{\overline{\mr{MS}}}

%%% note: COMMENT colors for arXiv submission
%
%\definecolor{Gray}{rgb}{0.5,0.5,0.5}
%\newcommand{\gry}{\color{Gray}}
%\definecolor{Red}{rgb}{0.9,0.0,0.0}
%\newcommand{\red}{\color{Red}}
%\definecolor{Blue}{rgb}{0.0,0.0,0.9}
%\newcommand{\blu}{\color{Blue}}
%\definecolor{Black}{rgb}{0.0,0.0,0.0}
%\newcommand{\bla}{\color{Black}}

%\hyphenation{bla-bla}

%%%%%%%%%%%%%%%%%%%%%%%%%%%%%%%%%%%%%%%%%%%%%%%%%%%%%%%%%%%%%%%%%%%%%%%%%%%%%%%%

\section{Introduction \label{sec:int}}

%%%%%%%%%%%%%%%%%%%%%%%%%%%%%%%%%%%%%%%%%%%%%%%%%%%%%%%%%%%%%%%%%%%%%%%%%%%%%%%%

In four spacetime dimensions (4D) staggered fermions offer a discretized version of four Dirac fermions or ``tastes'' \cite{Susskind:1976jm}.
But they are unequal -- two of them have positive and two have negative chirality \cite{Karsten:1980wd}.
Unlike flavor symmetry (which acts on different fields with the same fermion mass) the resulting taste symmetry is not exact, but broken by cut-off effects \cite{Sharatchandra:1981si,KlubergStern:1983dg,Golterman:1984cy,Smit:1986fn}.
There is a long stream of efforts which try to mitigate the effect of taste symmetry violation by adding evanescent terms%
\footnote{In practice, this is most conveniently done via ``smearing'' or ``gradient flow'', see Refs.~\cite{Blum:1996uf,Orginos:1998ue,Lagae:1998pe,Lepage:1998vj,Knechtli:2000ku}.}
to the staggered action \cite{Blum:1996uf,Orginos:1998ue,Lagae:1998pe,Lepage:1998vj,Knechtli:2000ku}
and/or by parameterizing the symmetry violation by a dedicated effective field theory description \cite{Kawamoto:1981hw,Lee:1999zxa}.

A fresh perspective on the problem was created by the discovery of ``minimally doubled'' fermions in 4D.
Both Karsten-Wilczek (KW) \cite{Karsten:1981gd,Wilczek:1987kw} and Bori\c{c}i-Creutz (BC) \cite{Creutz:2007af,Borici:2007kz} fermions are in this category.
Each one of these formulations encodes two species or ``tastes'' (with opposite chiralities), exactly the minimum required by the Nielsen-Ninomiya theorem \cite{Karsten:1980wd,Nielsen:1981xu,Nielsen:1980rz}.

Unfortunately, there is limited knowledge on the actual size of the taste breaking%
\footnote{Throughout, we use ``taste breaking'' as a shorthand for ``taste symmetry breaking''.}
effects of minimally doubled actions and how they fare relative to staggered taste breakings.
Staggered fermions are believed to show $O(a^2)$ cut-off effect, due to the (reduced) chiral symmetry
\cite{Sharatchandra:1981si,KlubergStern:1983dg,Golterman:1984cy,Smit:1986fn,Lepage:2011vr}.
Both KW and BC fermions have a (reduced) chiral symmetry \cite{Karsten:1981gd,Wilczek:1987kw,Creutz:2007af,Borici:2007kz},
and it has been argued that certain quantities have $O(a^2)$ cut-off effects \cite{Cichy:2008gk,Weber:2015oqf,Weber:2016dgo}.
But it is not clear (to us) whether these arguments cover \emph{all} quantities, in particular the taste splittings.

In two spacetime dimensions (2D) numerical investigations are much cheaper than in 4D.
But there is an important difference -- in 2D the staggered action is minimally doubled, too.
We see this as an opportunity to compare ``like with like'', that is three different discretization schemes which encode two species each.
Details of minimally doubled fermions in 2D (for instance w.r.t.\ topology and the free-field dispersion relation) have been investigated in
Refs.~\cite{Pernici:1994yj,Tiburzi:2010bm,Creutz:2010bm,Durr:2020yqa,Durr:2022mnz}.
Details of staggered fermions are found in
Refs.~\cite{Sharatchandra:1981si,KlubergStern:1983dg,Golterman:1984cy,Smit:1986fn,Smit:1987jh,Smit:1987fq,Laursen:1990ec,Hands:1990wc}.

We choose the simplest gauge group available, compact $U(1)$.
The resulting theory, known as ``Schwinger model'' \cite{Schwinger:1962tp,Lowenstein:1971fc}, resembles QCD in 4D, as it obeys an ``index theorem'' \cite{Ansourian:1977qe}.
This similarity implies that (even close to the continuum) one must be able to sample gauge field configurations with topological charge (or ``index'') $q\in\mathbb{Z}$ ergodically, without ``topology freezing''.
In the Schwinger model this problem has been solved long ago \cite{Smit:1987fq,Dilger:1992yn,Dilger:1994ma,Durr:2012te,Eichhorn:2021ccz}.

The remainder of this article is organized as follows.
Sec.~\ref{sec:sim} specifies the update algorithm and how a gauge configuration is assigned a topological charge $q\in\mathbb{Z}$.
Sec.~\ref{sec:ana} details how the taste splitting is determined from the low-lying eigenvalues.
Sec.~\ref{sec:lat} investigates how the splittings in a fixed physical volume depend on the lattice spacing $a$.
Sec.~\ref{sec:vol} checks, for fixed $a$, the volume dependence of these splittings.
Sec.~\ref{sec:sym} presents evidence that the splittings $\de$ obey a Symanzik power law $\de\propto a^p$ for $a\to0$.
Sec.~\ref{sec:con} contains our conclusions, with details of the topological charge determination shifted to App.~\ref{sec:app}.

%%%%%%%%%%%%%%%%%%%%%%%%%%%%%%%%%%%%%%%%%%%%%%%%%%%%%%%%%%%%%%%%%%%%%%%%%%%%%%%%

\section{Simulation setup and topological charge distributions \label{sec:sim}}

%%%%%%%%%%%%%%%%%%%%%%%%%%%%%%%%%%%%%%%%%%%%%%%%%%%%%%%%%%%%%%%%%%%%%%%%%%%%%%%%

We simulate the quenched Schwinger model with the Wilson gauge action
\beq
S[U]=\be\sum_x\big\{1-\mr{Re}[U_\dal(x)]\big\} \qquad\mbox{with}\qquad U_\dal(x)=U_1(x)U_2(x+\hat{1})U_1^*(x+\hat{2})U_2^*(x)
\label{def_swil}
\eeq
where $\hat{\mu}$ denotes $a$ times the unit vector in direction $\mu$, and $U^*$ is the complex conjugate of the unit-modulus variable $U$.
Alternatively, one might parameterize it as $U_\mu(x)=e^{\ri\vp_\mu(x)}$ with $\vp\in]-\pi,\pi]$,
and substitute $\mr{Re}[.]\rightarrow\cos[\vp_1(x)+\vp_2(x+\hat{1})-\vp_1(x+\hat{2})-\vp_2(x)]$ in (\ref{def_swil}).

\begin{table}[!b]\centering
\begin{tabular}{|l|cccccccc|}
\hline
$\beta$             &   3.2   &   5.0   &   7.2   &   12.8  &   20.0  &   28.8  &   51.2  &   80.0  \\
$L/a$               &    16   &    20   &    24   &    32   &    40   &    48   &    64   &    80   \\
$n_\mr{stout}$      &  0,1,3  &  0,1,3  &  0,1,3  &  0,1,3  &  0,1,3  &  0,1,3  &  0,1,3  &  0,1,3  \\
\hline
%%%
%%% grep elser_action splittings_2D_gen.bin_??.?_??_??_0_main
%%%
$s_\mr{wils}^{(0)}$ & 0.17625 & 0.10662 & 0.07230 & 0.03989 & 0.02533 & 0.01752 & 0.00981 & 0.00627 \\
%%%
%%% grep -A1 accrate splittings_2D_gen.bin_??.?_??_??_?_main | awk '{print $7}'
%%% remove finite-volume part
%%% obj=[...]; obj=reshape(obj,2,3,8); squeeze(mean(obj,2))
%%%      0.74996      0.73692      0.72915      0.72474      0.72565      0.72235      0.72146      0.72079
%%%    0.0014433    0.0014933       0.0015    0.0014967    0.0014967    0.0015167    0.0014833    0.0014767
%%%
$p_\mr{inst-hit}$   & 0.750(2) & 0.737(2) & 0.729(2) & 0.725(2) & 0.726(2) & 0.722(2) & 0.721(2) & 0.721(1) \\
\hline
\end{tabular}
\caption{\label{tab:cutoff}\sl
Overview of the ensembles used in the ``cut-off effect'' study; they implement constant physical volume through $(eL)^2=(L/a)^2/\be=80$.
For each choice of $(\be,L/a)$ three ensembles of 10\,000 configurations each are generated, to be used with 0, 1, 3 steps of $\rh=0.25$ stout smearing, respectively.
The analytic result $s_\mr{wils}^{(0)}$ for unsmeared Wilson glue is taken from Ref.~\cite{Elser:2001pe},
and $p_\mr{inst-hit}$ denotes the measured acceptance ratio in the instanton-hit update routine.}
\end{table}

We use multihit-Metropolis and overrelaxation sweeps in a $1\div4$ ratio, and augment these packages with
instanton hits (proposing a $q\to q\pm1$ update, which is accepted with a probability that tends to $1$ in the infinite-volume limit, see Tabs.~\ref{tab:cutoff} and \ref{tab:finvol})
and parity hits (proposing a $q\to-q$ update which is always accepted).
With these ingredients it is easy to generate well decorrelated configurations (see Ref.~\cite{Durr:2012te} for details).

\begin{table}\centering
\begin{tabular}{|l|cccccc|}
\hline
$\beta$           &   7.2    &   7.2    &   7.2    &   7.2    &   7.2    &   7.2    \\
$L/a$             &    16    &    20    &    24    &    32    &    40    &    48    \\
$n_\mr{stout}$    &    1     &    1     &    1     &    1     &    1     &    1     \\
\hline
$p_\mr{inst-hit}$ & 0.597(2) & 0.677(2) & 0.729(2) & 0.799(2) & 0.838(2) & 0.866(2) \\
\hline
\end{tabular}
\caption{\label{tab:finvol}\sl
Overview of the ensembles used in the ``finite volume'' study; each one contains 10\,000 configurations and is used after a single step of $\rh=0.25$ stout smearing.
In addition, the acceptance rate of the instanton hit update (see text) at the respective $(\be,L/a)$ is given.}
\end{table}

\begin{figure}[!tb]
\includegraphics[width=0.33\textwidth]{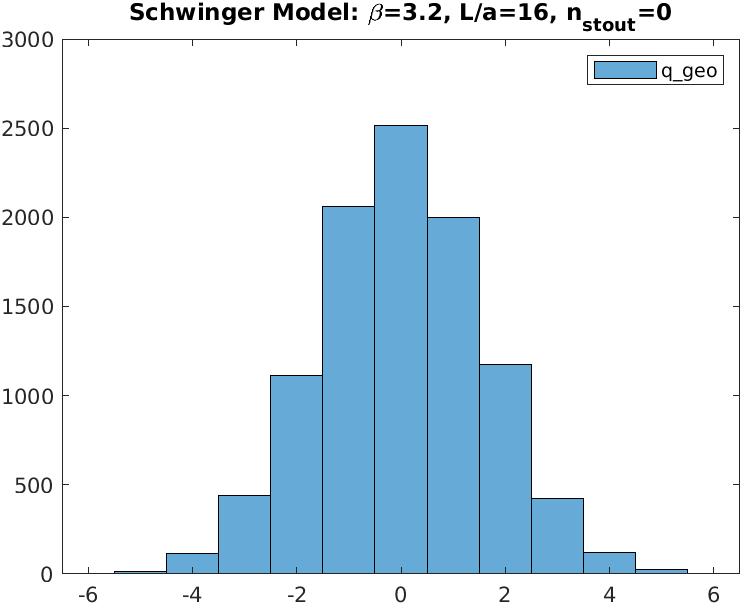}\hfill
\includegraphics[width=0.33\textwidth]{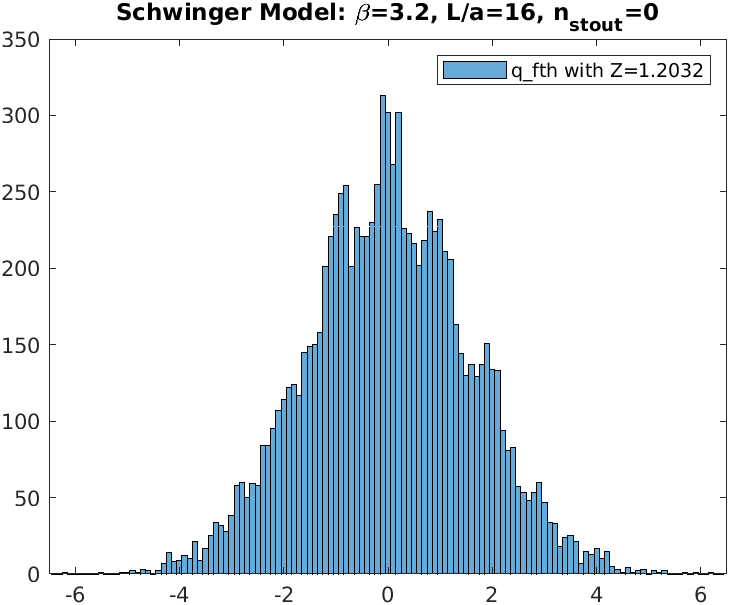}\hfill
\includegraphics[width=0.33\textwidth]{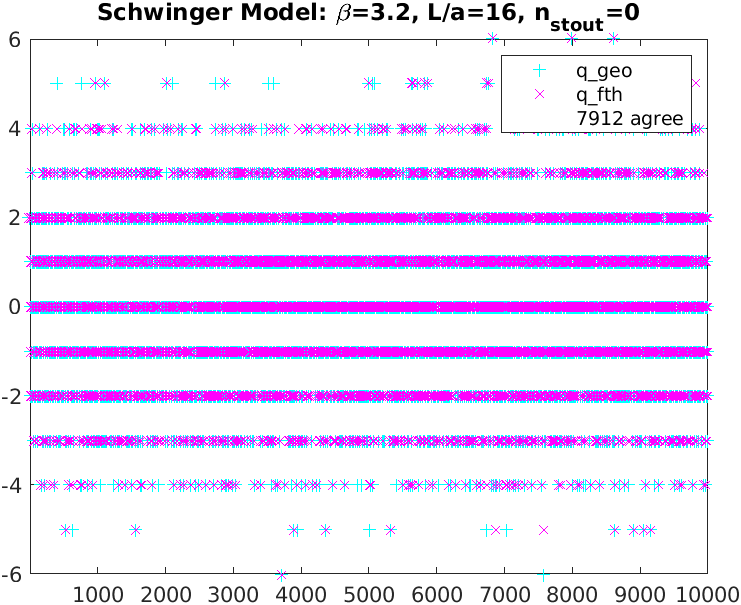}\\[2mm]
\includegraphics[width=0.33\textwidth]{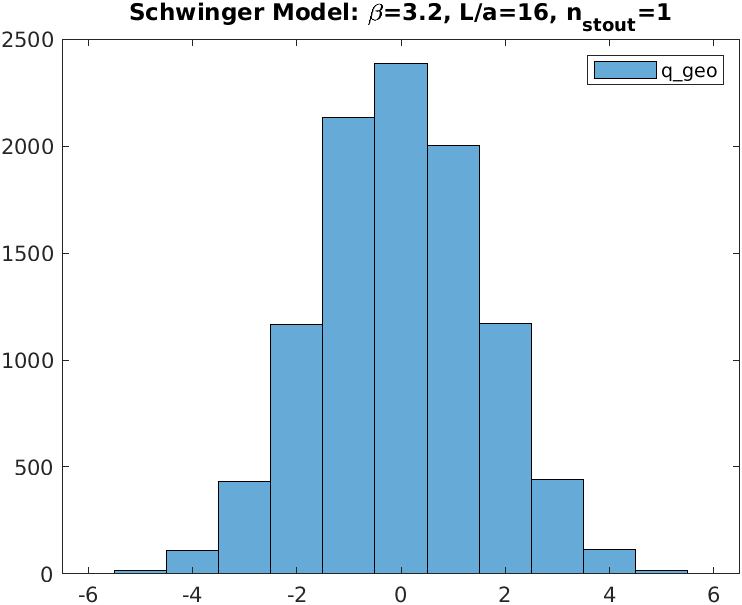}\hfill
\includegraphics[width=0.33\textwidth]{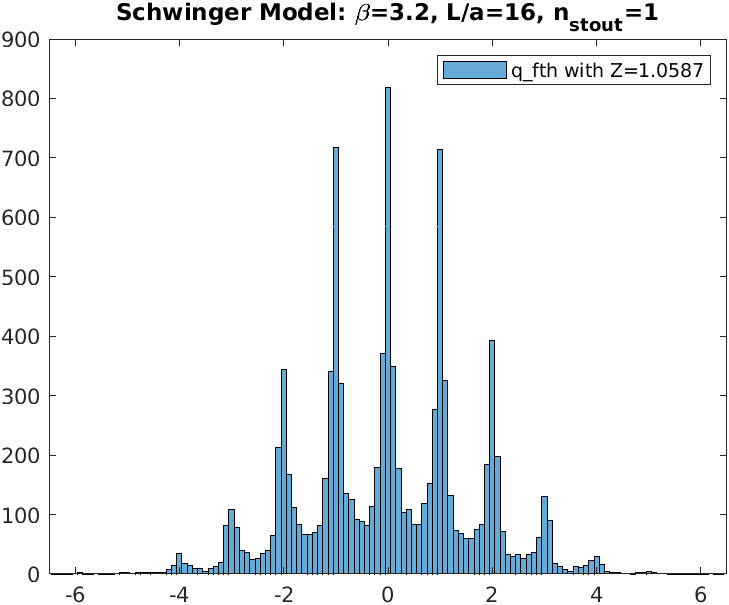}\hfill
\includegraphics[width=0.33\textwidth]{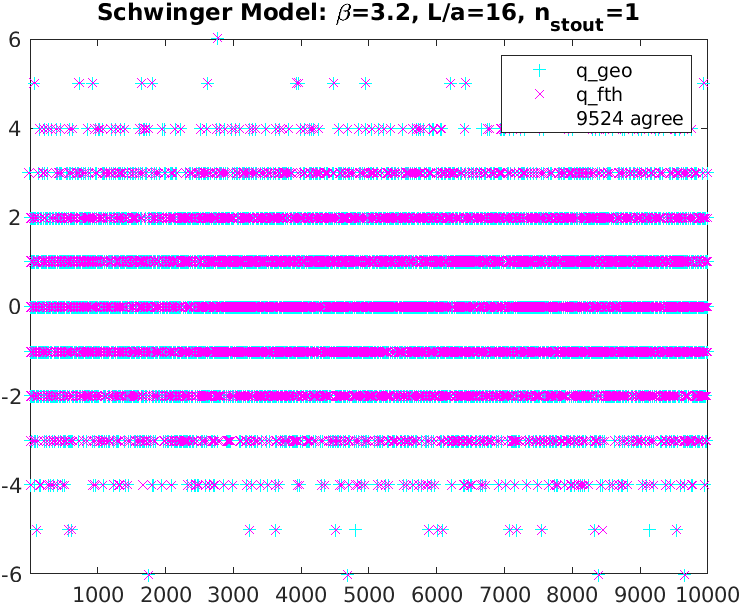}\\[2mm]
\includegraphics[width=0.33\textwidth]{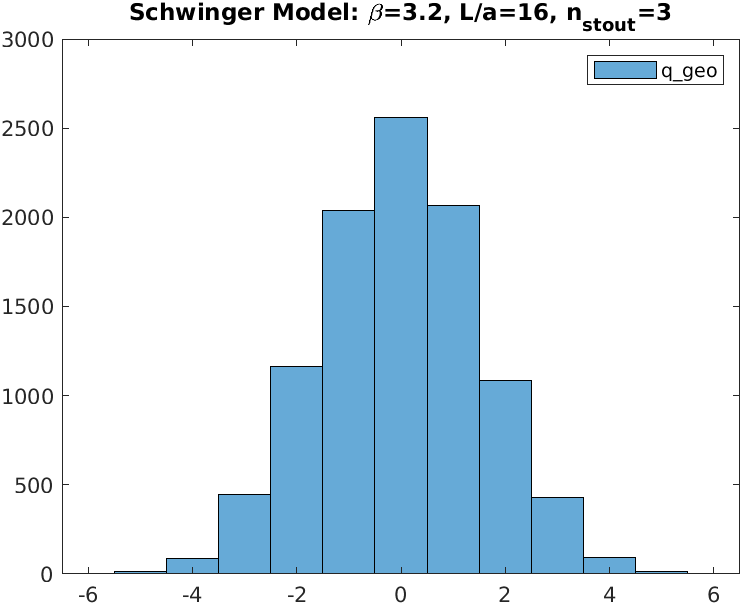}\hfill
\includegraphics[width=0.33\textwidth]{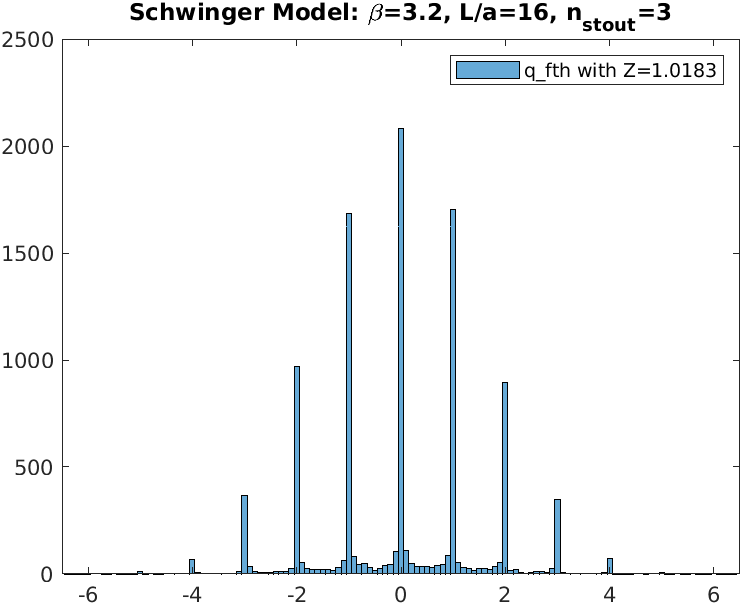}\hfill
\includegraphics[width=0.33\textwidth]{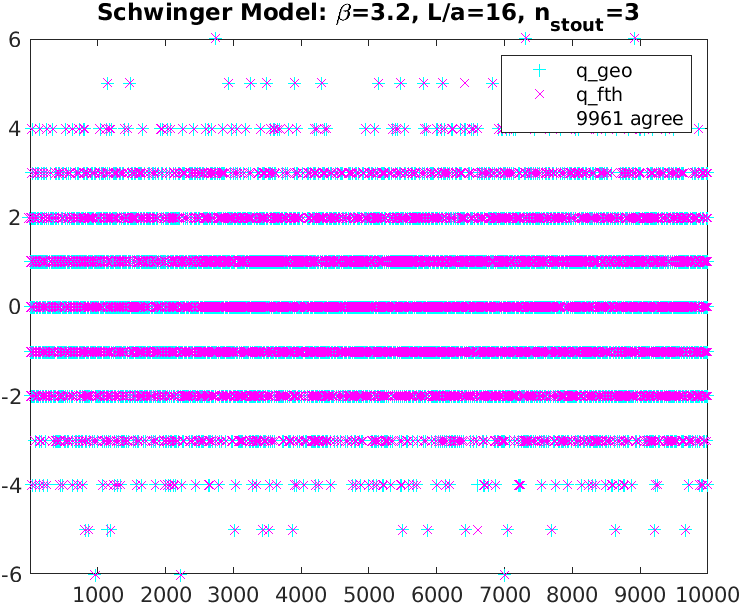}%
\vspace*{-2mm}
\caption{\label{fig:top}\sl
Histograms (for our coarsest ensembles) of the integer valued topological charge $q_\mr{geo}$ (left), and of the real valued $Zq_\mr{raw}$ (middle),
along with the resulting MC time histories of $q_\mr{geo}$ and $q_\mr{fth}$ after rounding (right)
with 0 (top), 1 (middle) and 3 (bottom) stout smearings.
The three rows refer to three different ensembles of 10\,000 configurations each at $(\be,L/a)=(3.2,16)$.}
\end{figure}

After the configurations are generated, we smear them with $n$ stout steps at $\rh=0.25$ \cite{Morningstar:2003gk},
and on the smeared backgrounds we evaluate the two topological charges \cite{Smit:1986fn,Smit:1987fq}
\bea
q_\mr{geo}^{(n)}\!&\!\equiv\!&\!\frac{1}{2\pi}\sum_x \mr{Im}\log U_\dal^{(n)}(x) = \frac{1}{2\pi}\sum_x     [\vp_1^{(n)}(x)+\vp_2^{(n)}(x+\hat{1})-\vp_1^{(n)}(x+\hat{2})-\vp_2^{(n)}(x)]_{{}_{]-\pi,\pi]}^{}}
\label{def_qgeo}
\\
q_\mr{raw}^{(n)}\!&\!\equiv\!&\!\frac{1}{2\pi}\sum_x \mr{Im}\,   U_\dal^{(n)}(x) = \frac{1}{2\pi}\sum_x \sin[\vp_1^{(n)}(x)+\vp_2^{(n)}(x+\hat{1})-\vp_1^{(n)}(x+\hat{2})-\vp_2^{(n)}(x)     ]
\label{def_qraw}
\eea
where $U_\dal^{(n)}(x)$ is the 4-link product in (\ref{def_swil}) after $n\in\{0,1,3\}$ smearing steps.
The ``geometric charge'' (\ref{def_qgeo}) is integer valued, while (\ref{def_qraw}) is not.
But the distribution of the latter is strongly peaked at near-integer values $\mathbb{Z}/Z$ with $Z>1$.
Hence, one defines the ``field-theoretic charge''
\beq
q_\mr{fth}^{(n)}=\mr{round}(Zq_\mr{raw}^{(n)})
\label{def_qfth}
\eeq
which is again integer-valued.
We determine $Z=Z(\be,n)$ non-perturbatively, with details given in App.~\ref{sec:app}.
In a particular run only one smearing level is realized, and the resulting configuration $U^{(n)}$ is assigned a topological charge $q^{(n)}$ if and
only if $q_\mr{geo}^{(n)}=q_\mr{fth}^{(n)}$ holds true%
\footnote{In principle also the integer-valued staggered topological charge $q_\mr{stag}^{(n)}$ might be evaluated,
see e.g.\ Ref.~\cite{Durr:2022mnz} for details in 2D and a guide to the literature. Due to the increased CPU demand, we refrain from doing so.}.
Fig.~\ref{fig:top} presents the histograms of $q_\mr{geo}^{(n)}$ and $q_\mr{fth}^{(n)}$ at our coarsest lattice spacing ($\be=3.2$) for the stouting levels $n\in\{0,1,3\}$.
The respective MC time histories look ergodic and the fraction of configurations without charge assignment diminishes quickly with increasing $n$ and/or increasing $\be$.

In a given run the smearing level $n\in\{0,1,3\}$ is kept fixed, and the matrices $\DS$, $\DKW$ and $\DBC$ are evaluated on the respective background $U^{(n)}$.
These operators refer to the Susskind ``staggered'' \cite{Susskind:1976jm}, Karsten-Wilczek \cite{Karsten:1981gd,Wilczek:1987kw}
and Bori\c{c}i-Creutz \cite{Creutz:2007af,Borici:2007kz} definitions of the massless Dirac matrix, respectively.
For each formulation the $16$ smallest eigenvalues $\ri\la$ on the positive%
\footnote{The eigenvalues come in $\pm\ri\la$ pairs, due to $\ep$-hermiticity or $\gaf$-hermiticity, respectively.}
imaginary axis are determined and the $\la>0$ are archived.
This way the eigenvalues of different formulations are statistically correlated, but the different smearing levels are not.

The Schwinger model (with any $\Nf$) is super-renormalizable, and the gauge coupling $e$ has mass-dimension one.
We exploit this feature to set the lattice spacing via $ae=1/\sqrt{\be}$.
This way it is easy to define simulation parameters which implement a continuum limit ``at constant physics'' (i.e.\ fixed box size and possibly pion mass in physical units), see Tab.~\ref{tab:cutoff}.
We produce an additional set of ensembles to investigate finite volume effects, see Tab.~\ref{tab:finvol} and pertinent comments in Sec.~\ref{sec:vol}.
In total this gives $3\times8+5=29$ ensembles for which we checked that the unsmeared plaquette agrees, within errors, with the analytic result of Ref.~\cite{Elser:2001pe}.

%%%%%%%%%%%%%%%%%%%%%%%%%%%%%%%%%%%%%%%%%%%%%%%%%%%%%%%%%%%%%%%%%%%%%%%%%%%%%%%%

\section{Analysis details on ``central ensemble'' \label{sec:ana}}

%%%%%%%%%%%%%%%%%%%%%%%%%%%%%%%%%%%%%%%%%%%%%%%%%%%%%%%%%%%%%%%%%%%%%%%%%%%%%%%%

The gauge ensemble with $(\be,L/a,n_\mr{stout})=(7.2,24,1)$ appears both in Tab.~\ref{tab:cutoff} and in Tab.~\ref{tab:finvol}.
We call it the ``central ensemble'' and use it to specify the details of our analysis.

In 2D the (rotated) eigenvalues $0<\la_1<...<\la_{16}$ of an operator $\DS$, $\DKW$ or $\DBC$ come in near-degenerate pairs (in 4D the staggered ones come in quartets \cite{Follana:2004sz,Durr:2004as}).
However, the details of the grouping depend on the absolute value of the topological charge of the background.
For $q=0$ the pairing is $\la_1\simeq\la_2,\la_3\simeq\la_4,...,\la_{15}\simeq\la_{16}$,
based on the fact that the intra-taste splitting is smaller than the physical eigenvalue splitting.
In this case $\sum_{j=1}^8\la_{2j}-\la_{2j-1}$ would be a suitable measure of the taste breaking effect on this background.
For $|q|=1$ things are different, since $\la_1$ is the archived part of the would-be zero mode pair $\pm\la_1$.
In this case $2\la_1+\sum_{j=1}^7\la_{2j+1}-\la_{2j}$ is an appropriate measure of the taste breaking effect (where we made sure that again eight splittings are taken into account).
Similarly, for $|q|=2$, one needs to keep in mind that $-\la_1,\la_1$ is one pair and $-\la_2,\la_2$ another one.
Hence, in this case we prefer to use $2\la_1+2\la_2+\sum_{j=1}^6\la_{2j+2}-\la_{2j+1}$, and so on.

% \begin{figure}[!tb]
% \includegraphics[width=0.49\textwidth]{gradflow_01.png}\hfill
% \includegraphics[width=0.49\textwidth]{gradflow_03.png}%
% \vspace*{-2mm}
% \caption{\label{fig:gradflow}\sl
% Evolution of the eigenvalues $a\la_j$ for $1 \leq j \leq 15$ (left) and taste splittings $a\de_1=2a\la_1$, $a\de_2=a\la_3-a\la_2$, etc.\ (right)
% of the staggered operator $a\DS$ on a single $(\be,L/a)=(7.2,24)$ configuration with $|q|=1$ under the gradient flow time $\ta/a^2$ (increased in steps of $0.05$).}
% \end{figure}

\begin{figure}[!tb]
\includegraphics[width=0.49\textwidth]{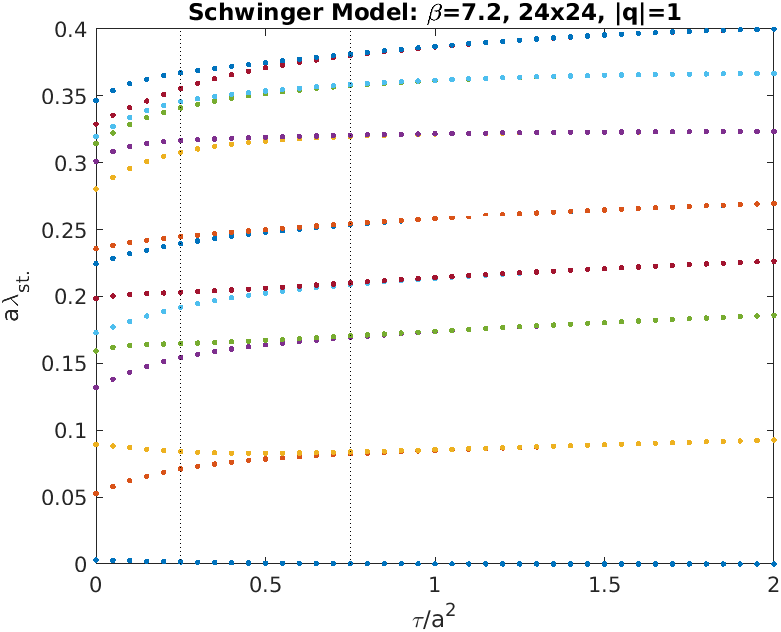}\hfill
\includegraphics[width=0.49\textwidth]{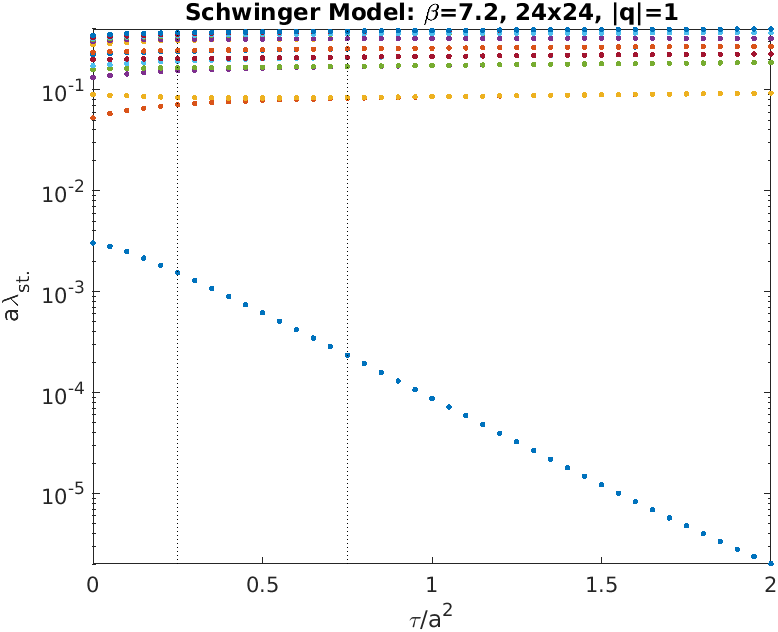}%
\vspace*{-2mm}
\caption{\label{fig:gradflow_lambdas}\sl
Eigenvalues $a\la_j$ ($1 \leq j \leq 15$) of $\DS/\ri$ on a $|q|=1$ configuration at $(\be,L/a)=(7.2,24)$
versus the gradient flow time $\ta/a^2$, in standard (left) and logarithmic (right) representation.}
\end{figure}

One may also consider each splitting separately, for instance
$\{2\la_1,2\la_2,\la_4-\la_3,...,\la_{14}-\la_{13}\}$ in the $|q|=2$ case, and this is what we shall do in the following.
Throughout, the operator $\DS$, $\DKW$ or $\DBC$ and the charge $q$ involve the same amount of link smearing, i.e.\ $n\in\{0,1,3\}$ steps of stout smearing \cite{Morningstar:2003gk} in both cases.

\begin{figure}[!tb]
\centering
\includegraphics[width=0.49\textwidth]{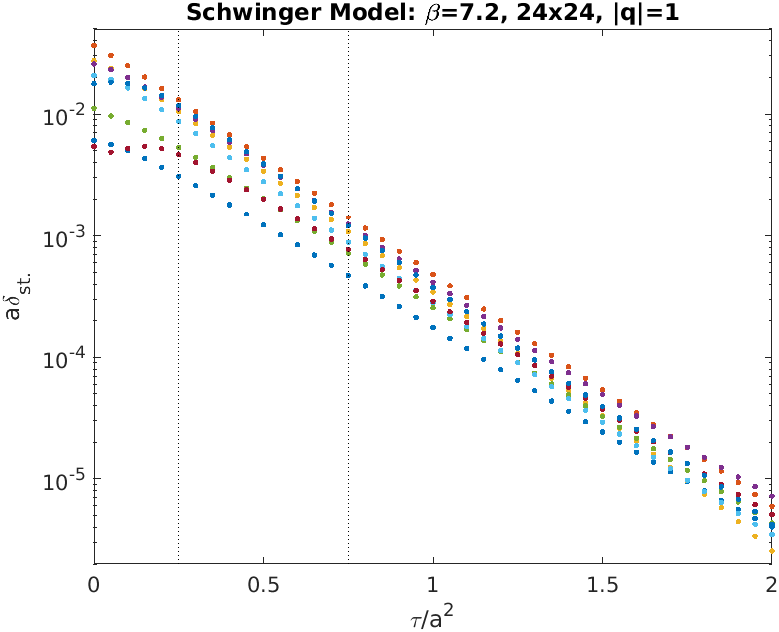}%
\vspace*{-2mm}
\caption{\label{fig:gradflow_deltas}\sl
Staggered taste splittings $a\de_1\equiv 2a\la_1, a\de_2\equiv a\la_3-a\la_2, ... ,a\de_8\equiv a\la_{15}-a\la_{14}$
(as appropriate for $|q|=1$) in logarithmic representation, derived from the data shown in Fig.~\ref{fig:gradflow_lambdas}.}
\end{figure}

Fig.~\ref{fig:gradflow_lambdas} displays, for a configuration with $|q|=1$, how the staggered eigenvalues $a\la_j$ evolve as a function of the gradient flow time $\ta/a^2$ \cite{Luscher:2010iy}.
The would-be zero mode $a\la_1$ (to be paired with its negative) goes to zero, all other modes form pairs that become gradually visible as the flow time increases.
Both the would-be zero mode and the remaining splittings (e.g.\ $a\de_2=a\la_3-a\la_2$) decrease \emph{exponentially} in the gradient flow time, see Fig.~\ref{fig:gradflow_deltas} and Ref.~\cite{Ammer:2022ksl}.
The flow times $\tau/a^2=0.25,0.75$ used in the main investigation (modulo discretization effects in $\ta$ \cite{Luscher:2010iy}) are marked with dotted vertical lines.
Hence, with one step of $\rh=0.25$ stout smearing at $\be=7.2$ the eigenvalue pairs are faintly visible, and with three steps they are easily identified.

\begin{figure}[!tb]
\includegraphics[width=0.49\textwidth]{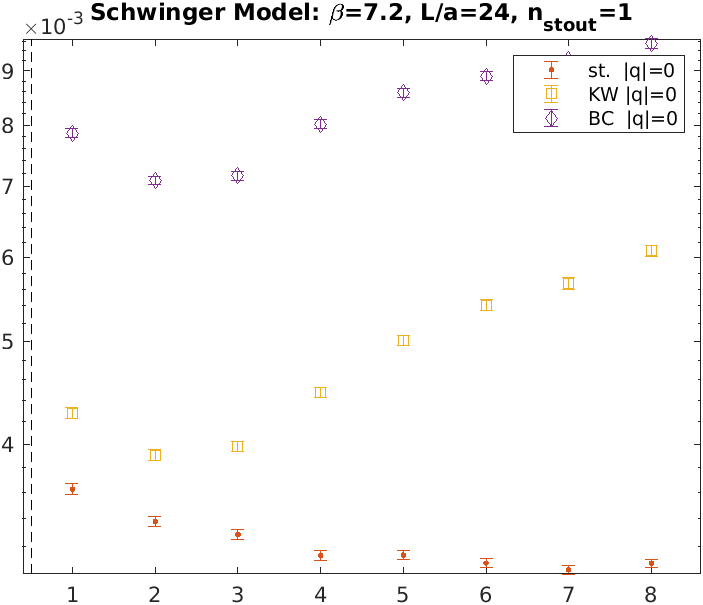}\hfill
\includegraphics[width=0.49\textwidth]{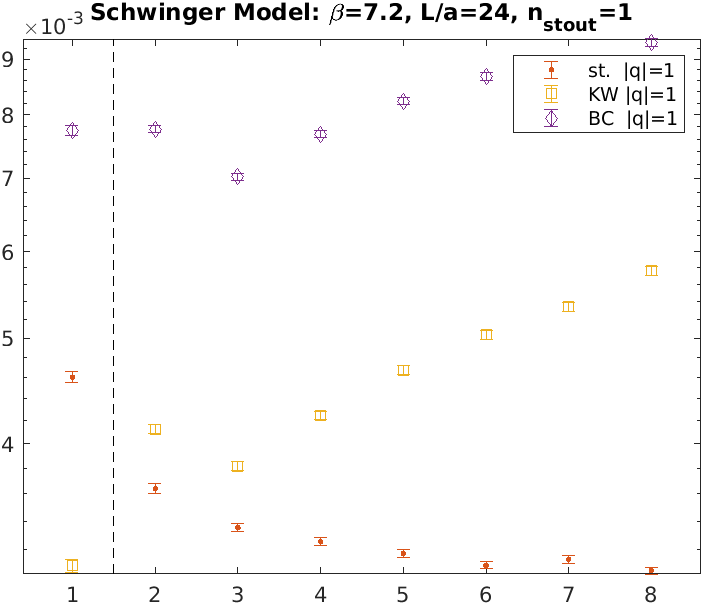}\\[2mm]
\includegraphics[width=0.49\textwidth]{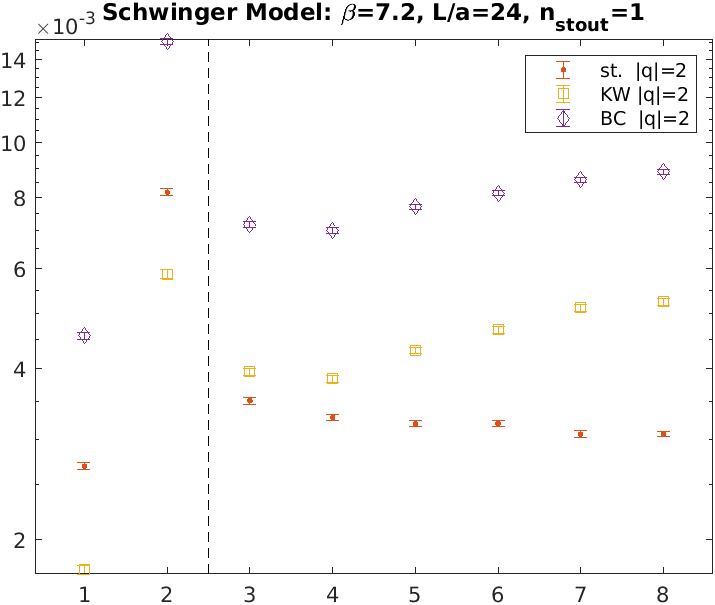}\hfill
\includegraphics[width=0.49\textwidth]{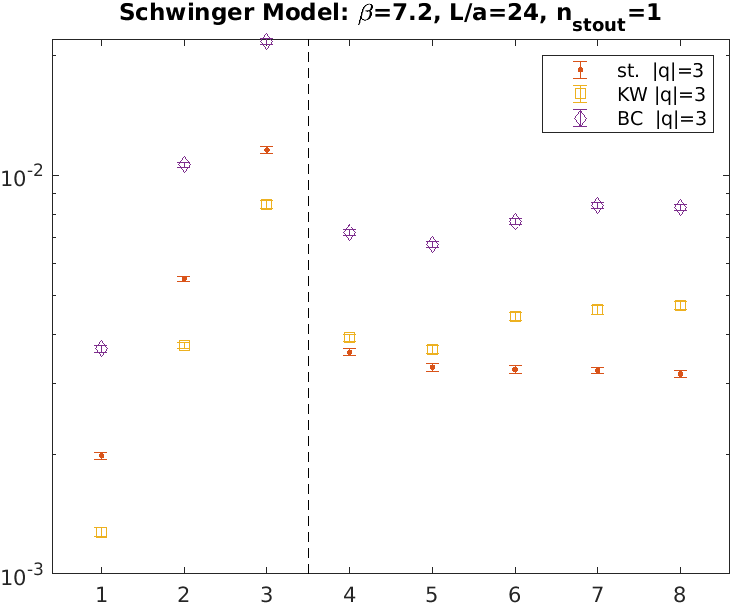}%
\vspace*{-2mm}
\caption{\label{fig:central}\sl
Taste-splittings $a\de_j$ ($1 \leq j \leq 8$) of the operators $\DS$, $\DKW$ and $\DBC$ on the ``central ensemble''.
The would-be zero mode splittings $a\de_1,...,a\de_{|q|}$ are separated from the non-topological splittings $a\de_{|q|+1},...$ by a dashed vertical line.
The four panels refer to $|q|=0,1,2,3$, respectively.
Both the operator and the charge measurement involve one step of stout smearing.}
\end{figure}

Fig.~\ref{fig:central} displays the taste splittings $a\de_j$ on our ``central ensemble'' $(\be,L/a,n_\mr{stout})=(7.2,24,1)$ in which all 10\,000 configurations are assigned%
\footnote{Respective numbers are $9524$ for $\be=3.2$, $9989$ for $\be=5.0$ and 10\,000 for $\be\geq7.2$ at $n_\mr{stout}=1$, cf.\ Tab.~\ref{tab:cutoff}.}
a topological charge.
The first panel shows the results on the $2697$ configurations with $q=0$; here staggered fermions feature the smallest taste breakings.
The second and third panels show the situation on the $4320$ and $2147$ configurations with $|q|=1$ and $|q|=2$, respectively.
The fourth panel repeats this for the $661$ configurations with $|q|=3$.
Throughout, the would-be zero mode splittings are separated by a vertical dashed line, and those of KW fermions seem particularly small.
The topological sectors $4\leq|q|\leq6$ hold too few ($151+20+4$) configurations for a statistical analysis.

In short, on the ``central ensemble'' no dramatic differences between the three fermion operators are observed.
For non-topological modes the staggered action shows the smallest taste-splittings, but for splittings linked to would-be zero modes the KW action performs better.
Hence, the question is whether this remains true as we vary the smearing level and/or the lattice spacing (Sec.~\ref{sec:lat}).
Also the impact of the box volume deserves a closer look (Sec.~\ref{sec:vol}).

%%%%%%%%%%%%%%%%%%%%%%%%%%%%%%%%%%%%%%%%%%%%%%%%%%%%%%%%%%%%%%%%%%%%%%%%%%%%%%%%

\section{Lattice spacing dependence \label{sec:lat}}

%%%%%%%%%%%%%%%%%%%%%%%%%%%%%%%%%%%%%%%%%%%%%%%%%%%%%%%%%%%%%%%%%%%%%%%%%%%%%%%%

As mentioned in Sec.~\ref{sec:sim}, we have the data to investigate how the unwanted taste splitting changes if the lattice spacing is varied at fixed (physical) box volume.

\begin{figure}[!p]
\includegraphics[width=0.49\textwidth]{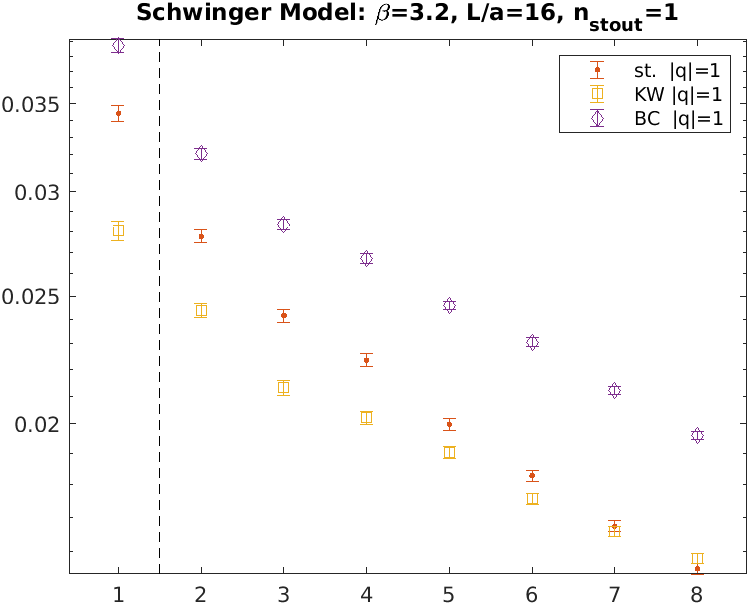}\hfill
\includegraphics[width=0.49\textwidth]{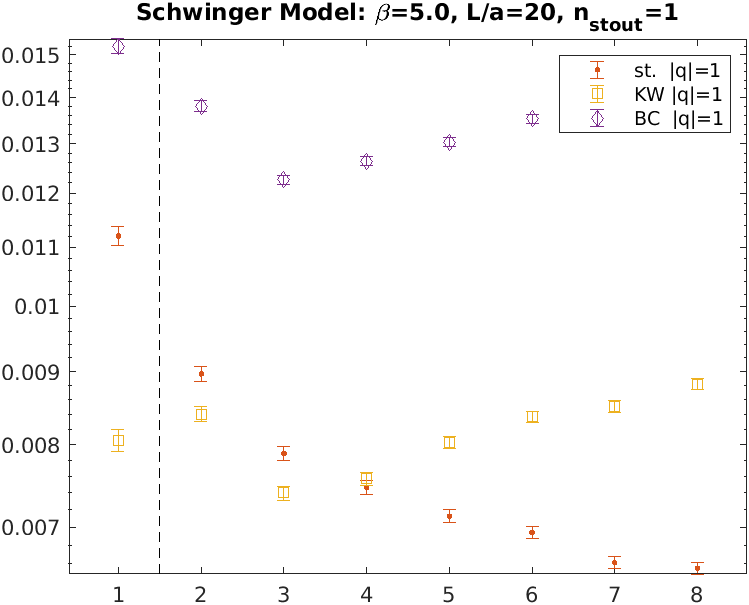}\\[2mm]
\includegraphics[width=0.49\textwidth]{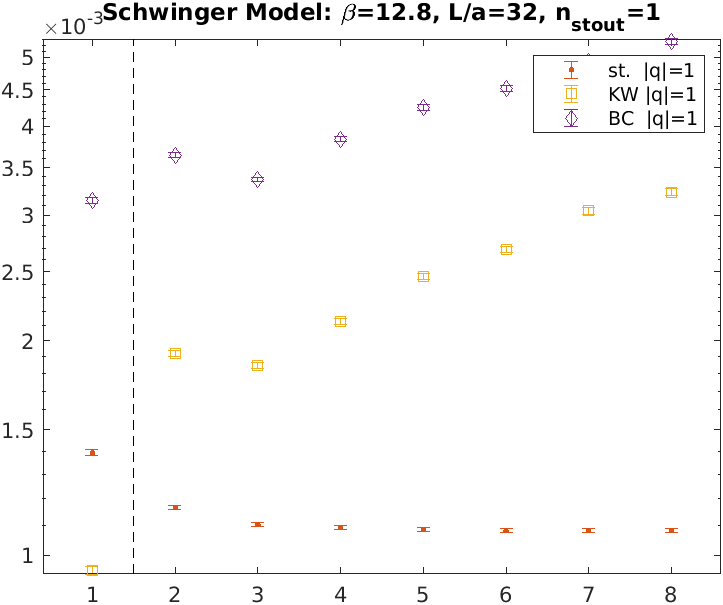}\hfill
\includegraphics[width=0.49\textwidth]{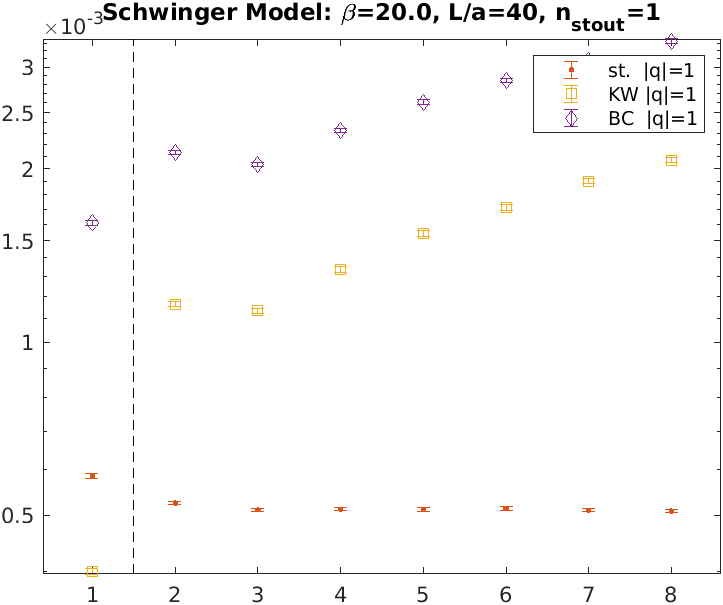}%
\vspace*{-2mm}
\caption{\label{fig:continuum1}\sl
Counterparts to the second ($|q|=1$) panel of Fig.~\ref{fig:central}, with coarser ($\be=3.2,5.0$) and finer ($\be=12.8,20.0$) lattice spacings, respectively.
The smearing level is $n_\mr{stout}=1$ throughout.}
\vspace*{6mm}
\includegraphics[width=0.49\textwidth]{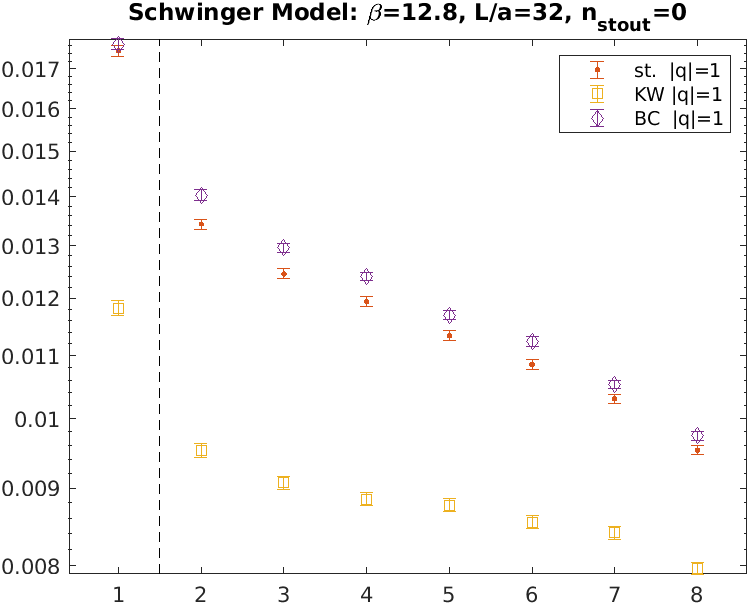}\hfill
\includegraphics[width=0.49\textwidth]{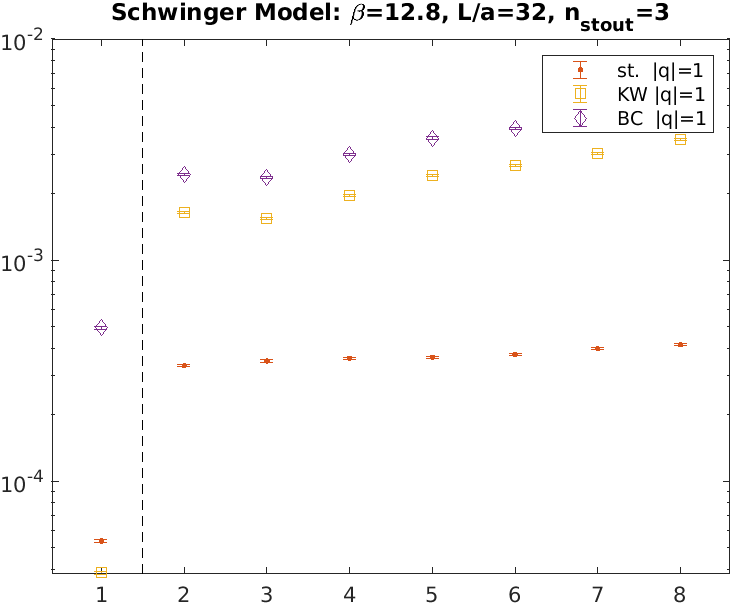}%
\vspace*{-2mm}
\caption{\label{fig:continuum2}\sl
Counterparts to the third ($\be=12.8$) panel of Fig.~\ref{fig:continuum1}, with $n_\mr{stout}=0,3$, respectively.}
\end{figure}

\begin{figure}[!p]
\includegraphics[width=0.49\textwidth]{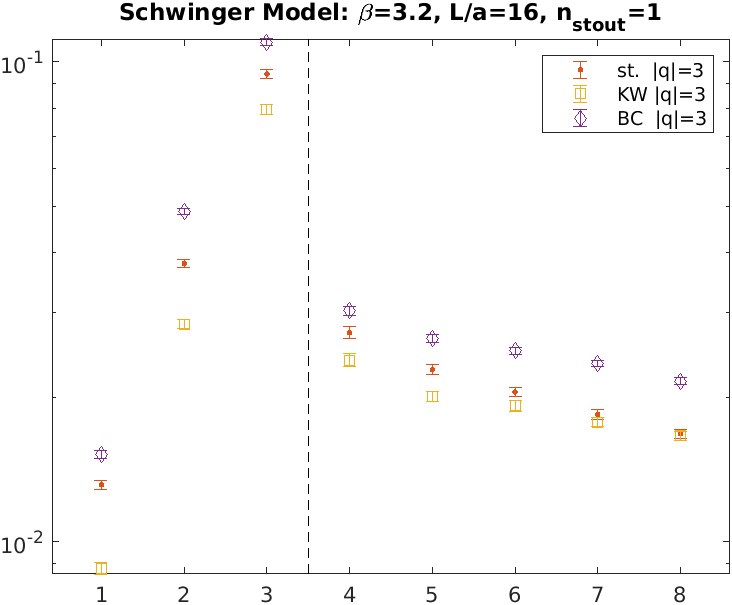}\hfill
\includegraphics[width=0.49\textwidth]{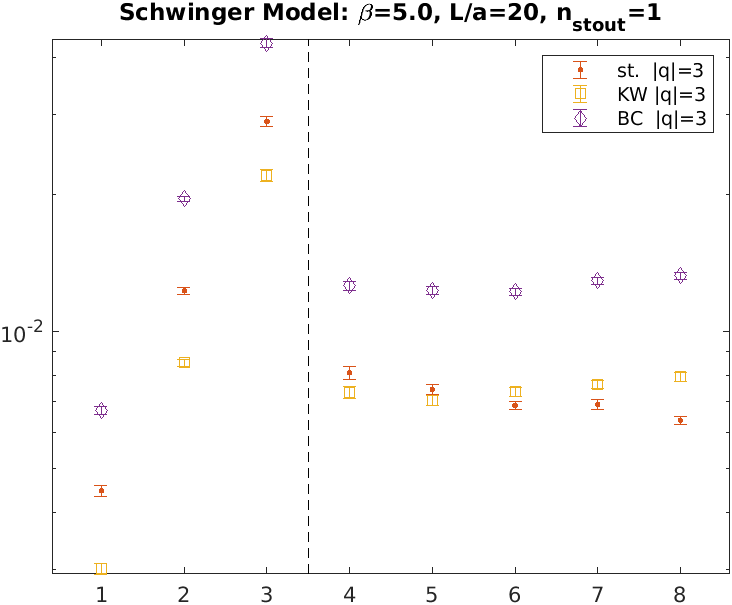}\\[2mm]
\includegraphics[width=0.49\textwidth]{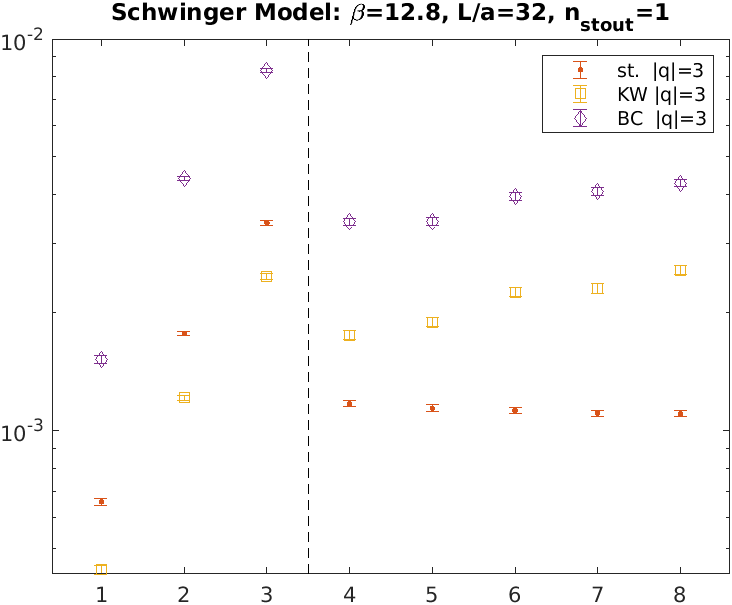}\hfill
\includegraphics[width=0.49\textwidth]{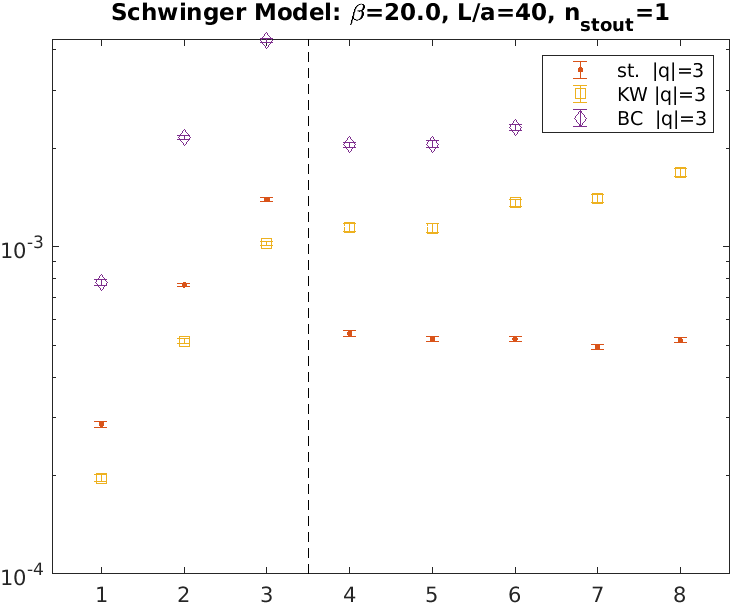}%
\vspace*{-2mm}
\caption{\label{fig:continuum3}\sl
Counterparts to the fourth ($|q|=3$) panel of Fig.~\ref{fig:central}, with coarser ($\be=3.2,5.0$) and finer ($\be=12.8,20.0$) lattice spacings, respectively.
The smearing level is $n_\mr{stout}=1$ throughout.}
\vspace*{6mm}
\includegraphics[width=0.49\textwidth]{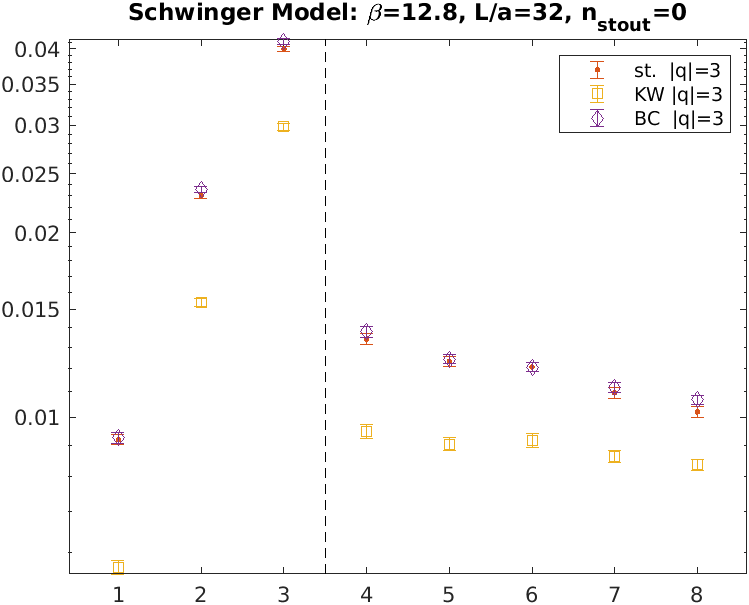}\hfill
\includegraphics[width=0.49\textwidth]{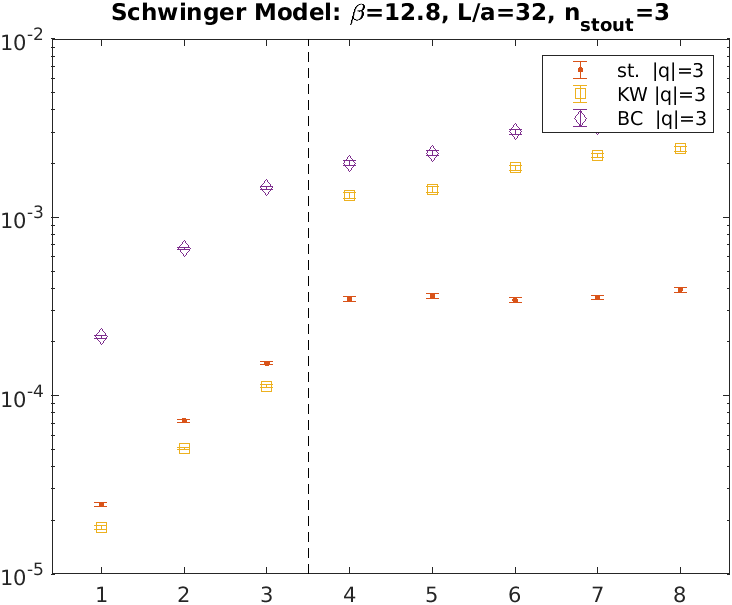}%
\vspace*{-2mm}
\caption{\label{fig:continuum4}\sl
Counterparts to the third ($\be=12.8$) panel of Fig.~\ref{fig:continuum3}, with $n_\mr{stout}=0,3$, respectively.}
\end{figure}

Let us recap the second panel ($|q|=1$) of Fig.~\ref{fig:central}, which was for $(\be,L/a,n_\mr{stout})=(7.2,24,1)$.
For the non-topological splittings ($j\geq2$) there is a hierarchy $\de_\mr{st}<\de_\mr{KW}<\de_\mr{BC}$.
For the would-be zero mode splitting ($j=1$), there is an inversion, since we find $\de_\mr{KW}<\de_\mr{st}<\de_\mr{BC}$.

The counterparts to this panel at coarser and finer lattice spacings are shown in Fig.~\ref{fig:continuum1}, still with $n_\mr{stout}=1$.
The most prominent change is a change in the $y$-axis label.
At $\be\in\{3.2,5.0\}$ the situation resembles the one of the ``central ensemble'', except that $\de_j^\mr{KW}$ wins for a few more $j$.
At weaker coupling, i.e.\ for $\be\in\{12.8,20.0\}$, the staggered splittings get progressively smaller (starting with the higher $j$) than the splittings of KW or BC fermions.
However, the would-be zero mode (to the left of the dashed vertical line) seems to be exempt from this rule; here the KW splitting remains smaller than any other splitting.
We have three finer lattice spacings $(\be=28.8,51.2,80.0$), but the situation remains similar to the one shown in the fourth panel.
Hence, close enough to the continuum, there is a difference between would-be zero modes and non-topological modes.
For large enough $\be$ the staggered action wins the contest for non-topological modes.
By contrast, for the first would-be zero mode the KW fermion features the smallest splitting at all accessible lattice spacings.

On may ask whether it makes a difference if we choose fewer ($n_\mr{stout}=0$) or more ($n_\mr{stout}=3$) smearing steps.
Fig.~\ref{fig:continuum2} presents these alternatives to the third panel ($\be=12.8$) of Fig.~\ref{fig:continuum1}.
In fact, the $n_\mr{stout}=0$ panel resembles the first panel in the previous figure, and the $n_\mr{stout}=3$ panel is similar to the fourth panel in the previous figure.
It seems that increasing/decreasing the smearing level acts a bit like%
\footnote{Of course, this statement is to be taken with a grain of salt; one should stay away from ``oversmearing'',
as this will cause lots of near-degeneracies among the eigenvalues, similar to the free-field case where excessive degeneracies emerge.
We maintain that all our smearing levels, $n_\mr{stout}=0,1,3$, represent \emph{mild} smearings,
since the integrated flow times $\ta/a^2=0.0,0.25,0.75$ yield diffusion lengths in lattice units $\sqrt{8\ta}/a\simeq0.0,1.41,2.45$, respectively \cite{Luscher:2010iy}.
We expect that the effects of ``oversmearing'' would show up in the regime $\ta/a^2\gg1$.}
increasing/decreasing $\be$.

In Fig.~\ref{fig:continuum3} the eigenvalue splittings of the $|q|=3$ configurations at $\be=3.2,5.0,12.8,20.0$ with $n_\mr{stout}=1$ are shown.
Together with the fourth panel of Fig.~\ref{fig:central} they constitute a ``line of constant physics''.
The qualitative difference between non-topological modes ($j=1,2,3$) and would-be zero modes ($j\geq4$) is quite obvious.
The three lattice spacings not shown $(\be=28.8,51.2,80.0$) feature a situation similar to one shown in the last panel.
Hence, for the non-topological modes the staggered action wins the contest at large enough $\be$, followed by KW and BC fermions.
For the would-be zero modes, on the other hand, the KW splitting is always smaller than the staggered one, and the latter fares better than the BC splitting.

In Fig.~\ref{fig:continuum4} the third panel ($\be=12.8$) of Fig.~\ref{fig:continuum3} is confronted with its siblings at $n_\mr{stout}=0$ and $n_\mr{stout}=3$.
Again the left panel resembles the first panel of the previous figure, and the right panel resembles the fourth panel of the previous figure.
Overall it seems that an increased smearing level gives a ``preview'' of a larger $\be$ (with the caveat mentioned above).

In short, varying the cut-off (at fixed $n_\mr{stout}=1$) confirms the different behavior of would-be zero modes and non-topological modes.
For the latter category the staggered action yields asymptotically smaller splittings than either minimally doubled action.
On the other hand, for the former category the KW action features the smallest splittings at all $(\be,n_\mr{stout})$ combinations explored.
We shall speculate on possible reasons for this observation in Sec.~\ref{sec:sym}.

%%%%%%%%%%%%%%%%%%%%%%%%%%%%%%%%%%%%%%%%%%%%%%%%%%%%%%%%%%%%%%%%%%%%%%%%%%%%%%%%

\section{Finite volume effects \label{sec:vol}}

%%%%%%%%%%%%%%%%%%%%%%%%%%%%%%%%%%%%%%%%%%%%%%%%%%%%%%%%%%%%%%%%%%%%%%%%%%%%%%%%

As mentioned in Sec.~\ref{sec:sim}, we have the data needed to investigate how the unwanted taste splitting changes if the (physical) box volume is varied at fixed lattice spacing.

% \begin{figure}[!tb]
% \includegraphics[width=0.49\textwidth]{fig_07.2_16_1_10.png}\hfill
% \includegraphics[width=0.49\textwidth]{fig_07.2_20_1_10.png}\\[2mm]
% \includegraphics[width=0.49\textwidth]{fig_07.2_32_1_10.png}\hfill
% \includegraphics[width=0.49\textwidth]{fig_07.2_40_1_10.png}%
% \vspace*{-2mm}
% \caption{\label{fig:volume1}\sl
% Counterparts to the second ($|q|=1$) panel of Fig.~\ref{fig:central}, with smaller ($L/a=16,20$) and larger ($L/a=32,40$) box size, respectively.
% Throughout $(\be,n_\mr{stout})=(7.2,1)$ is kept fixed.}
% \end{figure}
%
% The data in Fig.~\ref{fig:volume1} extend the second panel ($|q|=1$) of Fig.~\ref{fig:central} towards smaller and larger box sizes, keeping $(\be,n_\mr{stout})=(7.2,1)$ fixed.
% It seems the box volume does not impact the hierarchy in the eigenvalue splittings (except that the data in the last panel are a bit shaky, since in the $L/a=32$ run the number of
% $|q|=1$ configurations becomes small).

\begin{figure}[!tb]
\includegraphics[height=0.4\textwidth]{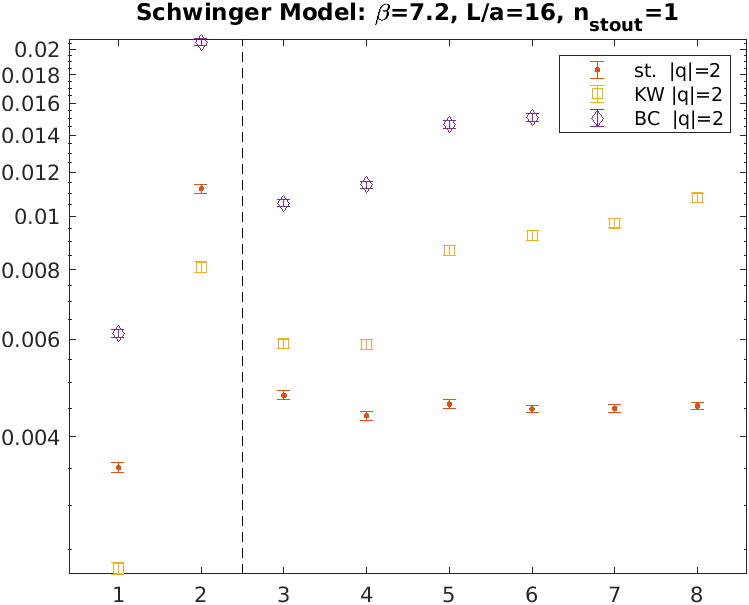}\hfill
\includegraphics[height=0.4\textwidth]{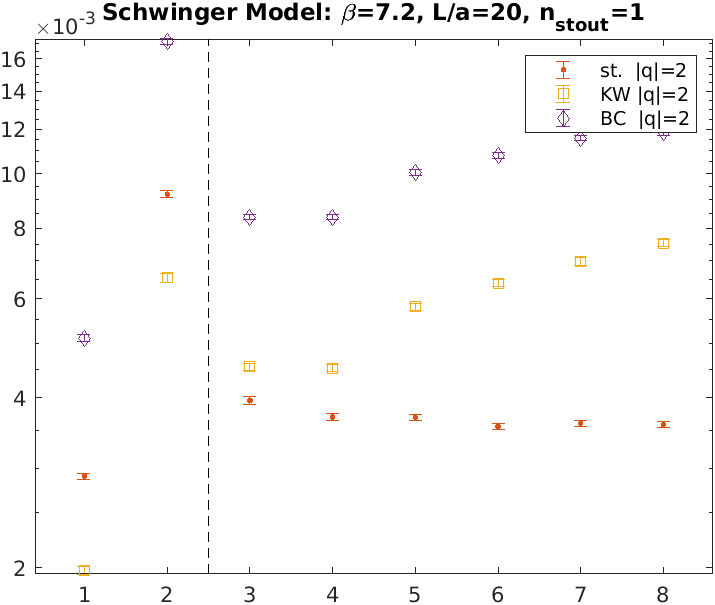}\\[2mm]
\includegraphics[height=0.4\textwidth]{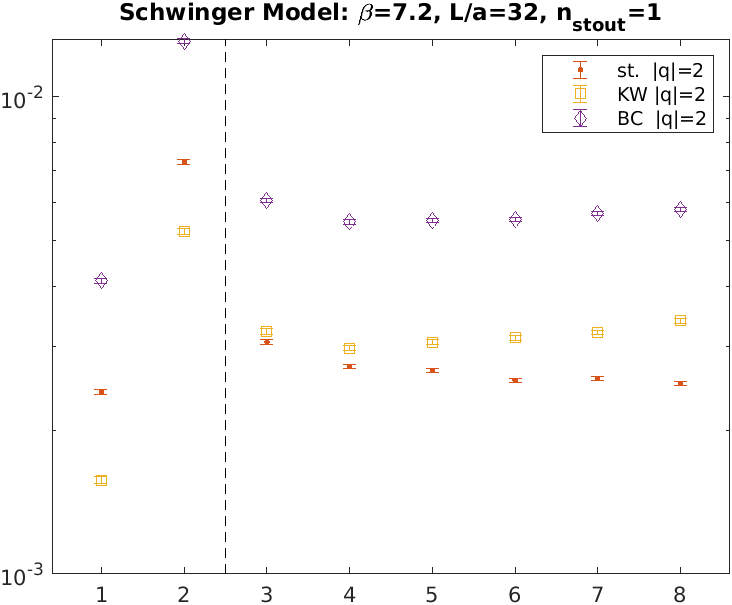}\hfill
\includegraphics[height=0.4\textwidth]{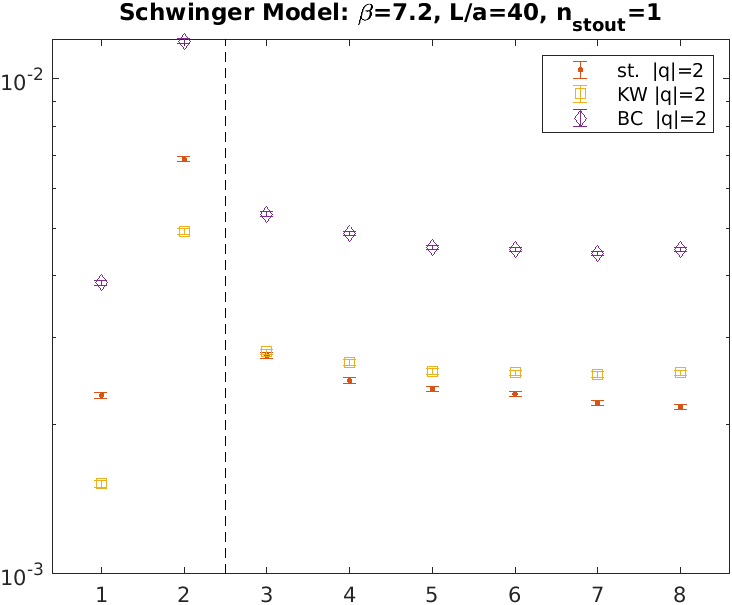}%
\vspace*{-2mm}
\caption{\label{fig:volume2}\sl
Counterparts to the third ($|q|=2$) panel of Fig.~\ref{fig:central}, with smaller ($L/a=16,20$) and larger ($L/a=32,40$) box size, respectively.
Throughout $(\be,n_\mr{stout})=(7.2,1)$ is kept fixed.}
\end{figure}

The data in Fig.~\ref{fig:volume2} extend the third panel ($|q|=2$) of Fig.~\ref{fig:central} towards smaller and larger box sizes, keeping $(\be,n_\mr{stout})=(7.2,1)$ fixed.
Evidently, the box volume impacts the overall scale, but it does not affect the hierarchy among the eigenvalue splittings.
In particular the distinction between would-be zero modes and non-topological modes holds in the sense that for these parameters the hierarchies
$\de_\mr{KW}<\de_\mr{st}<\de_\mr{BC}$ for would-be zero modes and $\de_\mr{st}<\de_\mr{KW}<\de_\mr{BC}$ for non-topological modes hold in all volumes.

% \begin{figure}[!tb]
% \includegraphics[width=0.49\textwidth]{fig_07.2_16_1_12.png}\hfill
% \includegraphics[width=0.49\textwidth]{fig_07.2_20_1_12.png}\\[2mm]
% \includegraphics[width=0.49\textwidth]{fig_07.2_32_1_12.png}\hfill
% \includegraphics[width=0.49\textwidth]{fig_07.2_40_1_12.png}%
% \vspace*{-2mm}
% \caption{\label{fig:volume3}\sl
% Counterparts to the fourth ($|q|=3$) panel of Fig.~\ref{fig:central}, with smaller ($L/a=16,20$) and larger ($L/a=32,40$) box size, respectively.
% Throughout $(\be,n_\mr{stout})=(7.2,1)$ is kept fixed.}
% \end{figure}
% 
% The data in Fig.~\ref{fig:volume3} extend the fourth panel ($|q|=3$) of Fig.~\ref{fig:central} towards smaller and larger box sizes, keeping $(\be,n_\mr{stout})=(7.2,1)$ fixed.
% Again the box volume does not seem to impact the overall structure of the data; the splittings of the three physical would-be zero modes are clearly set apart from the splittings of the non-topological modes.

\begin{figure}[!tb]
\centering
\includegraphics[height=0.39\textwidth]{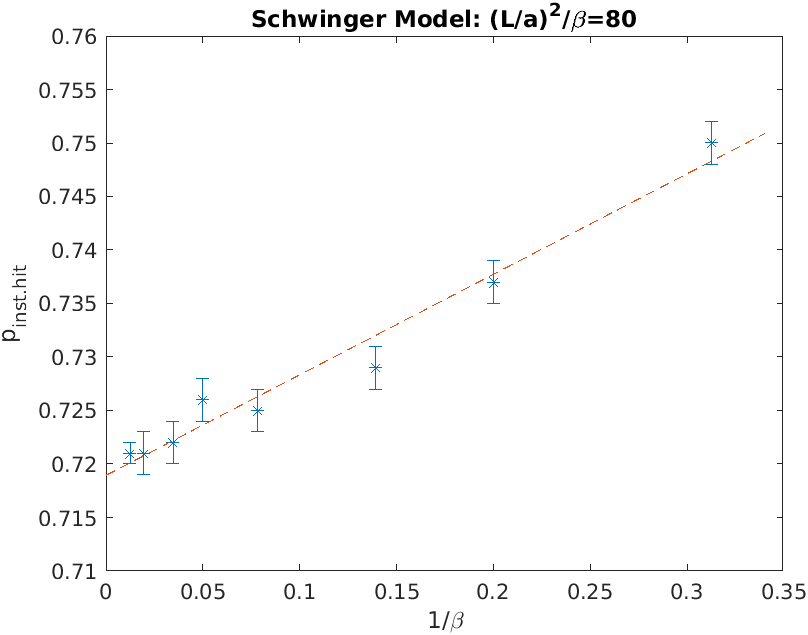}\hfill
\includegraphics[height=0.39\textwidth]{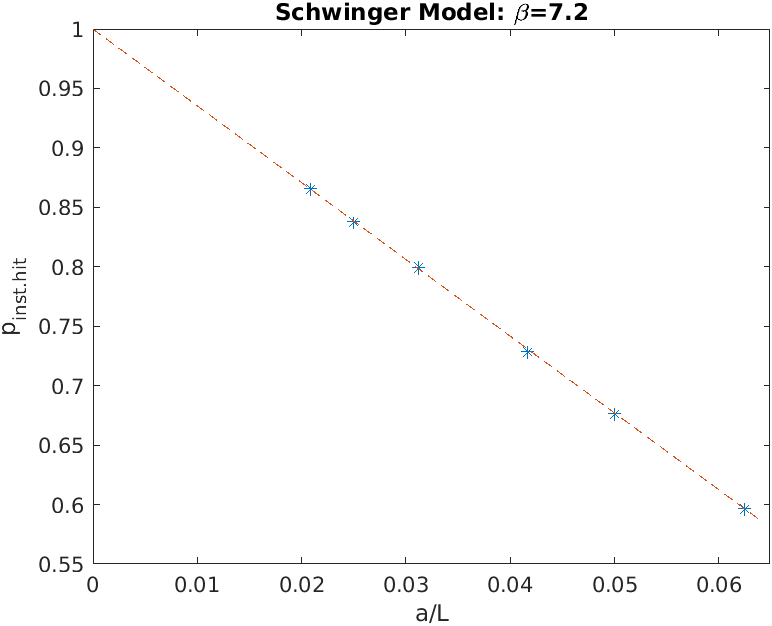}%
\vspace*{-2mm}
\caption{\label{fig:instantonhit}\sl
Instanton hit acceptance ratio $p_\mr{inst-hit}$ versus $1/\be$ in a fixed physical volume (left and Tab.~\ref{tab:cutoff}),
and versus $a/L$ at fixed $\be$ (right and Tab.~\ref{tab:finvol}), along with linear fits.}
\end{figure}

We conclude that the data presented in Sec.~\ref{sec:lat} have been collected in a large enough physical volume to avoid relevant finite-volume effects on the splittings.
Hence the question whether we understand the observed lattice spacing dependence is well warranted.
As an aside we find that the \emph{instanton hit acceptance ratio} depends in the first place on the \emph{physical volume};
the data in Fig.~\ref{fig:instantonhit} (right) suggest that $p_\mr{inst-hit}\to1$ holds in the limit $L\to\infty$ for any $\be$.

%%%%%%%%%%%%%%%%%%%%%%%%%%%%%%%%%%%%%%%%%%%%%%%%%%%%%%%%%%%%%%%%%%%%%%%%%%%%%%%%

\section{Symanzik scaling of taste splittings \label{sec:sym}}

%%%%%%%%%%%%%%%%%%%%%%%%%%%%%%%%%%%%%%%%%%%%%%%%%%%%%%%%%%%%%%%%%%%%%%%%%%%%%%%%

So far we found that it makes a difference whether a given eigenvalue belongs to a would-be zero mode or a non-topological mode.
For would-be zero modes KW fermions win%
\footnote{This solidifies and extends an observation reported in Ref.~\cite{Durr:2022mnz}.}
the contest; their splittings are smaller than those of staggered or BC fermions at any $(\be,n_\mr{stout})$ explored.
For non-topological modes staggered fermions yield the smallest intra-taste splittings if $n_\mr{stout}\geq1$.
These findings were presented in Figs.\,\ref{fig:central}--\ref{fig:volume2}.

The next step is to check whether a given splitting shows asymptotic Symanzik scaling, and, if so, to determine its universality class (i.e.\ the power $p$ in $\de \propto a^p$)
\cite{Symanzik:1983dc,Symanzik:1983gh,Curci:1983an,Luscher:1984xn}.
For this purpose one needs to collect all data pertinent to a given Dirac operator, smearing level and splitting type (i.e.\ treating would-be zero modes and non-topological modes separately),
and investigate their dependence on the lattice spacing $a\propto\be^{-1/2}$.
For staggered fermions standard reasoning suggests that the taste splittings in physical units (with mass dimension one) scale as $\de_i\propto a^2$ (tantamount to
$a\de_j\propto a^3$ in lattice units) for all $j$, unless the measurement operator reduces the power.
For KW and BC fermions things are more involved, since their chiral symmetry groups%
\footnote{See Ref.~\cite{Weber:2023kth} for a thorough analysis of the symmetry groups of KW and BC fermions.}
are smaller.
Still, there are results in the literature which suggest leading $O(a^2)$ cut-off effects for these fermion formulations \cite{Cichy:2008gk,Weber:2015oqf,Weber:2016dgo}, but in some cases%
\footnote{For instance Ref.~\cite{Durr:2020yqa} finds that the real part of the BC dispersion relation has leading cut-off effects $O([am]^2)$,
while the imaginary part has $O(am)$ cut-off effects.}
results are inconclusive.

%%%%%%%%%%

\begin{figure}[!p]
\includegraphics[width=0.49\textwidth]{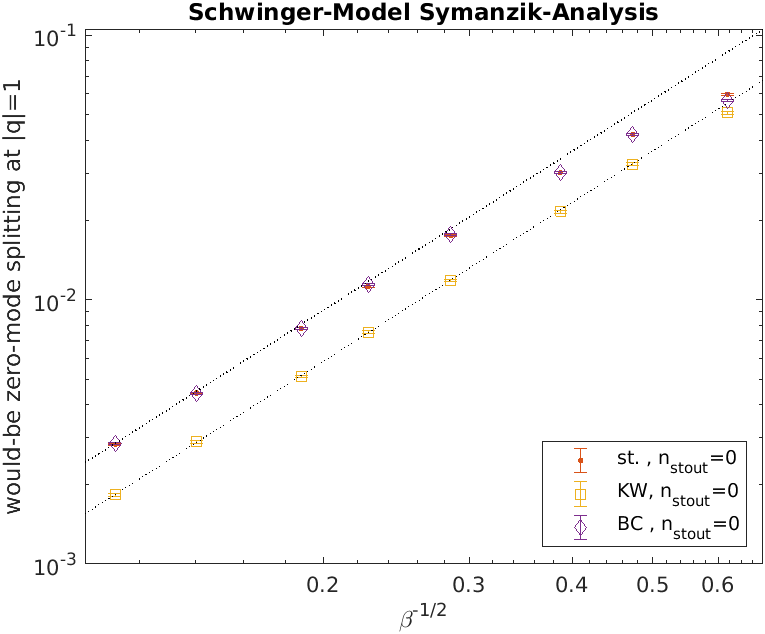}\hfill
\includegraphics[width=0.49\textwidth]{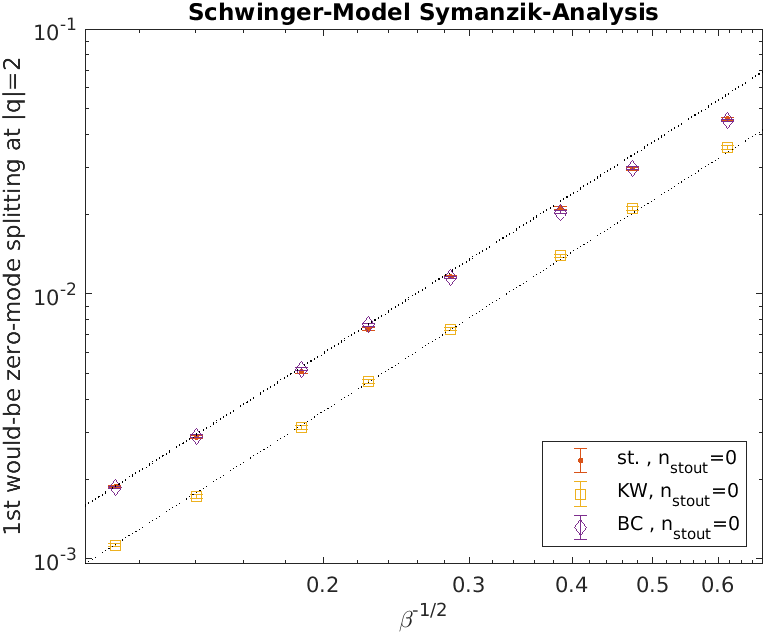}\\[2mm]
\includegraphics[width=0.49\textwidth]{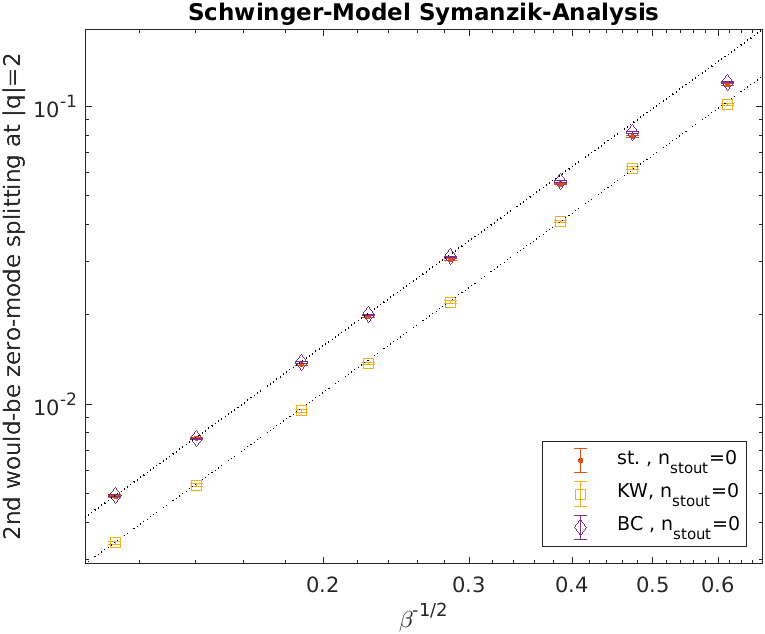}\hfill
\includegraphics[width=0.49\textwidth]{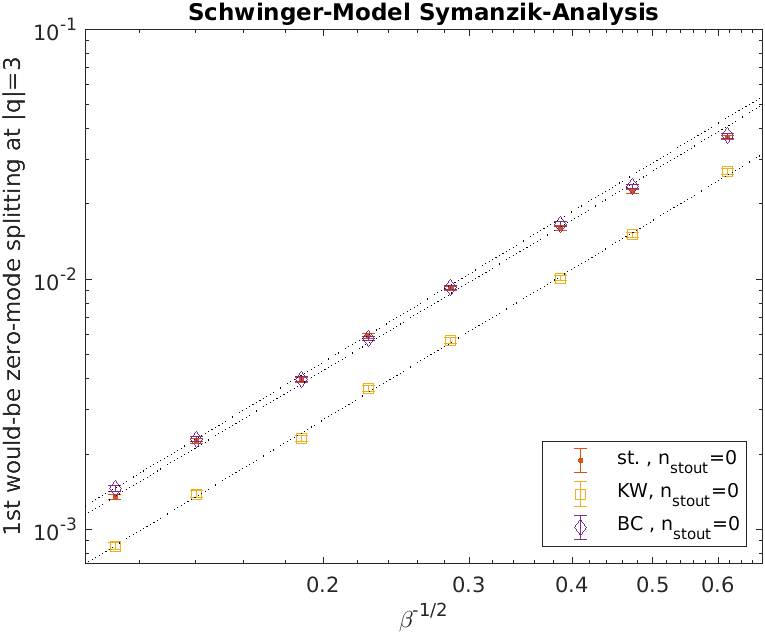}\\[2mm]
\includegraphics[width=0.49\textwidth]{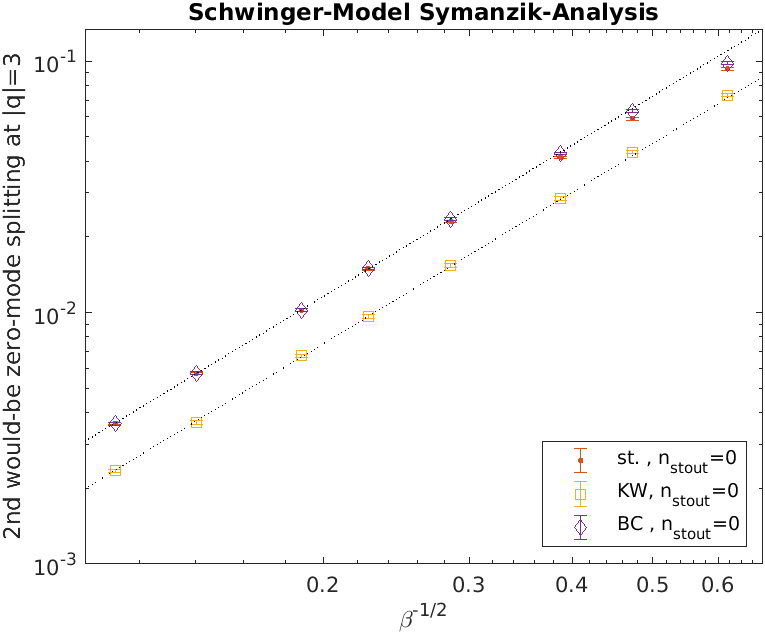}\hfill
\includegraphics[width=0.49\textwidth]{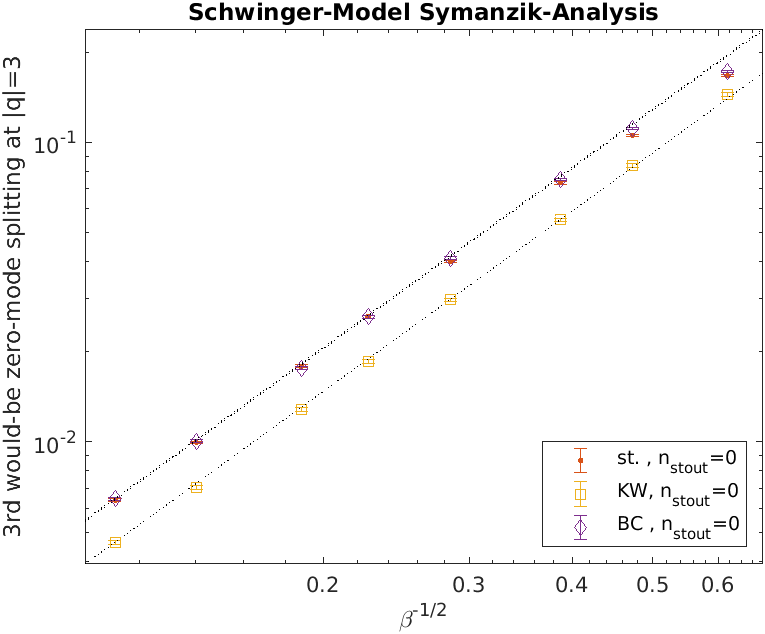}%
\caption{\label{fig:symanzik_wouldbe_n0}\sl
All would-be zero mode splittings $a\de_j$ at $|q|=1,2,3$ versus $a$ for $n_\mr{stout}=0$.
For each operator the dotted line is a power law $a\de \propto a^2$, going through the leftmost datapoint.}
\end{figure}

\begin{figure}[!p]
\includegraphics[width=0.49\textwidth]{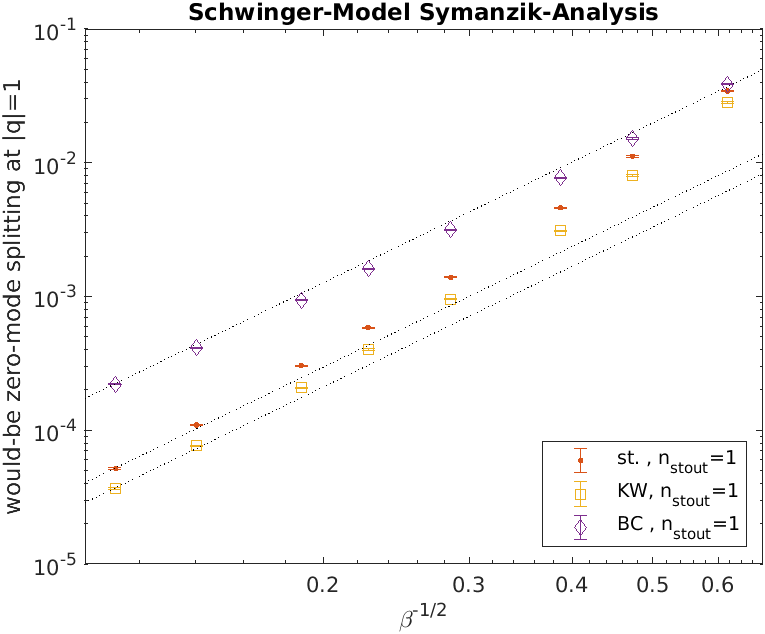}\hfill
\includegraphics[width=0.49\textwidth]{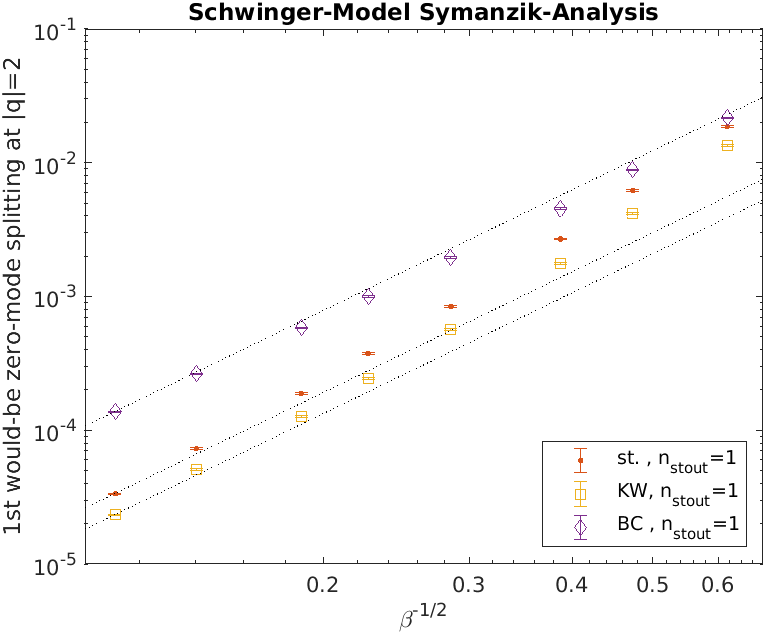}\\[2mm]
\includegraphics[width=0.49\textwidth]{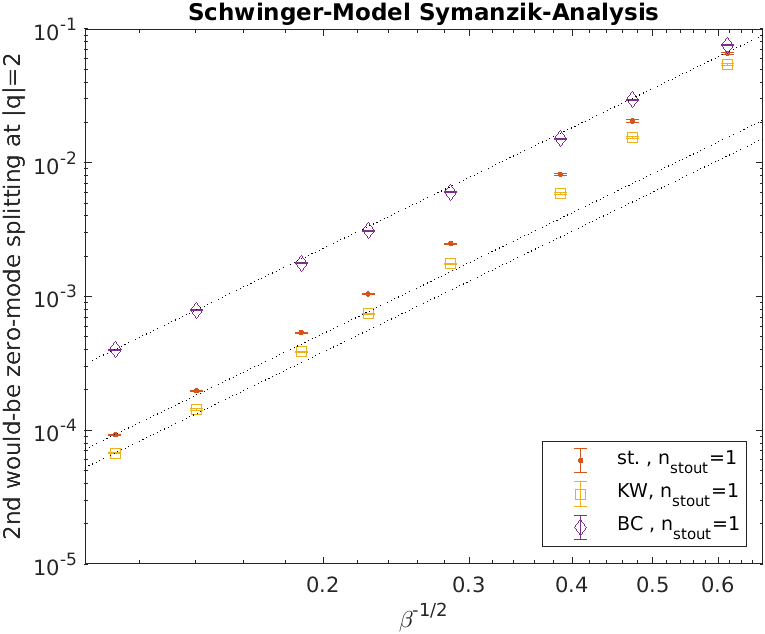}\hfill
\includegraphics[width=0.49\textwidth]{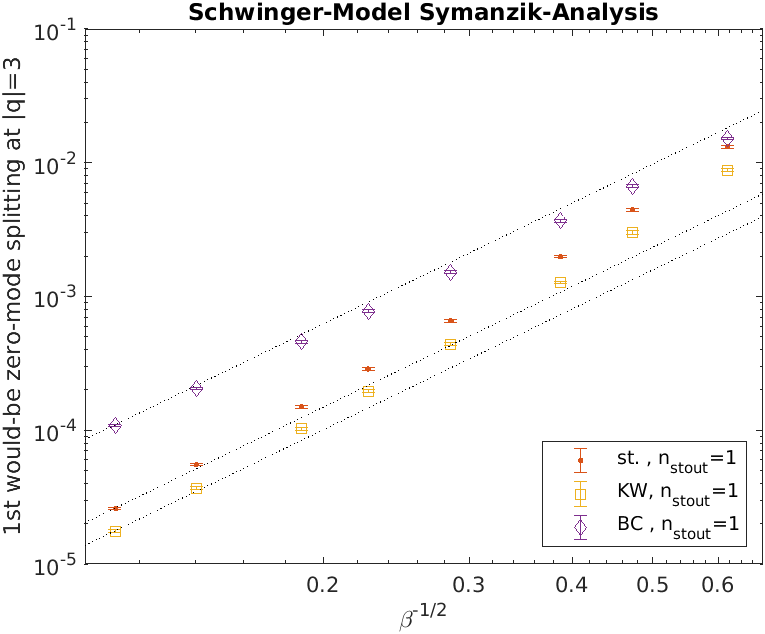}\\[2mm]
\includegraphics[width=0.49\textwidth]{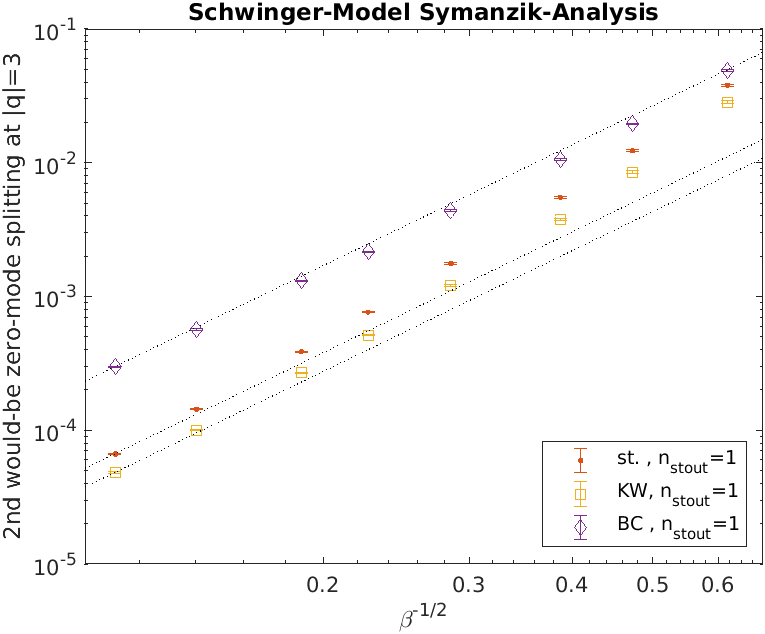}\hfill
\includegraphics[width=0.49\textwidth]{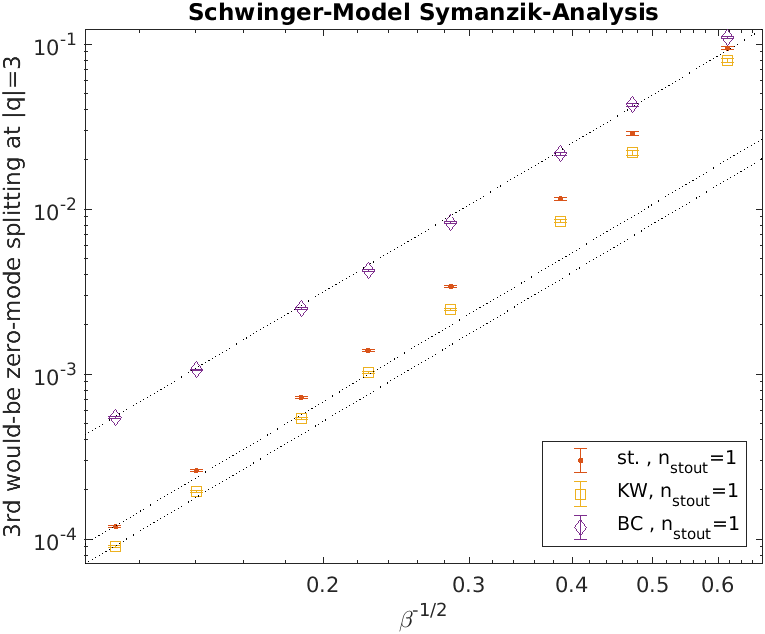}%
\caption{\label{fig:symanzik_wouldbe_n1}\sl
Same as Fig.~\ref{fig:symanzik_wouldbe_n0} but for $n_\mr{stout}=1$.
For each operator the dotted line is a power law $a\de \propto a^3$ (i.e.\ a higher power than in Fig.~\ref{fig:symanzik_wouldbe_n0}), going through the leftmost datapoint.}
\end{figure}

\begin{figure}[!p]
\includegraphics[width=0.49\textwidth]{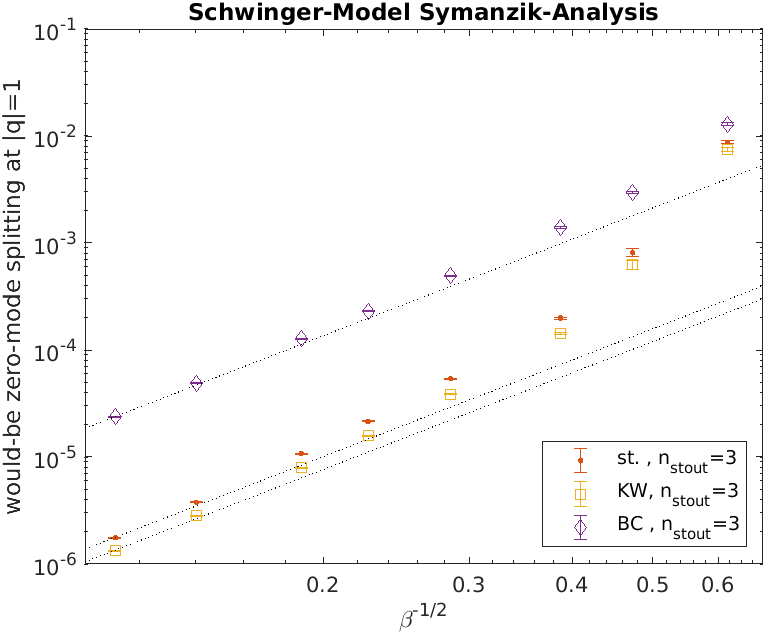}\hfill
\includegraphics[width=0.49\textwidth]{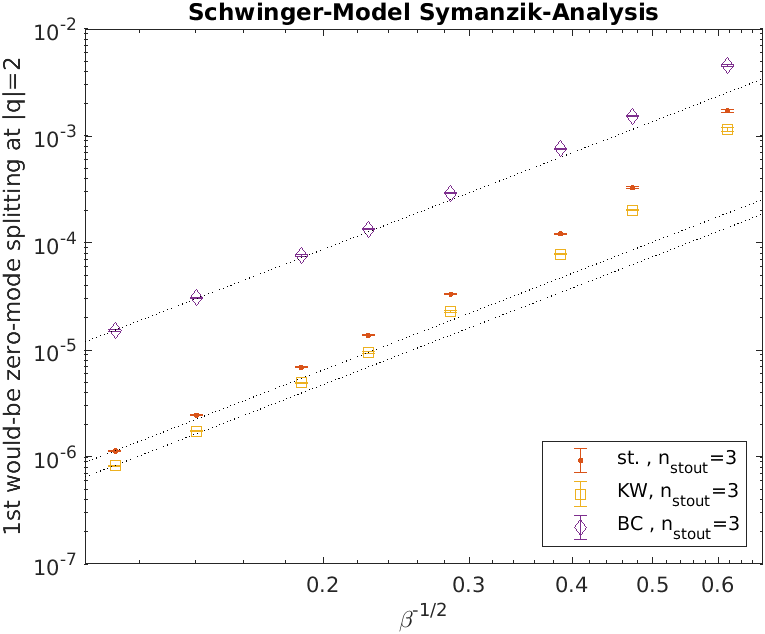}\\[2mm]
\includegraphics[width=0.49\textwidth]{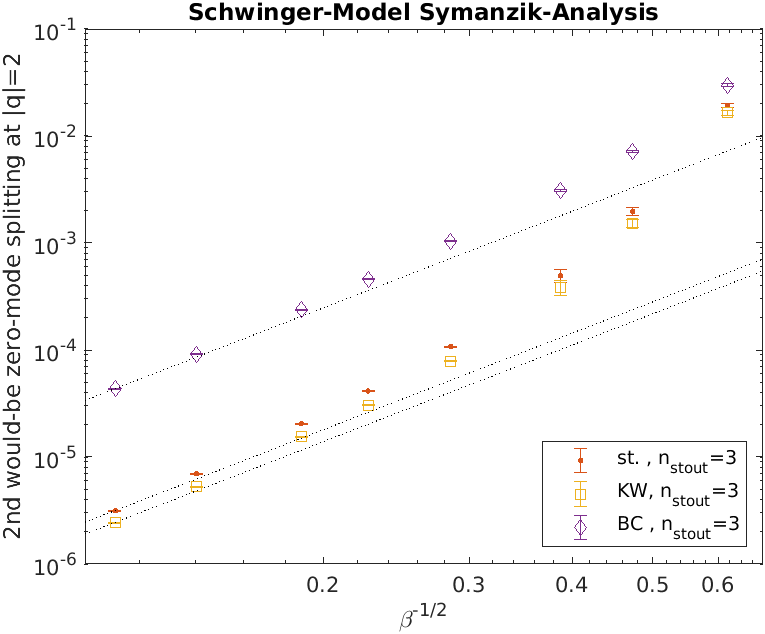}\hfill
\includegraphics[width=0.49\textwidth]{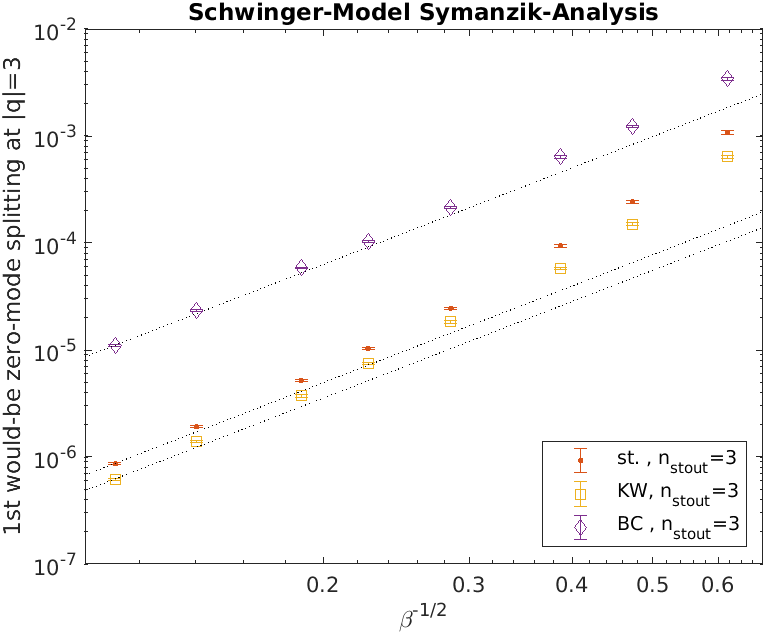}\\[2mm]
\includegraphics[width=0.49\textwidth]{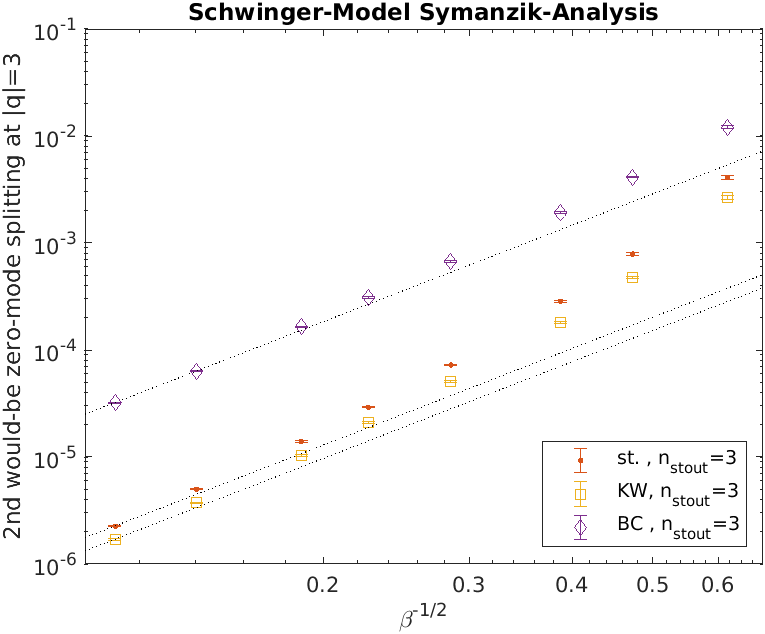}\hfill
\includegraphics[width=0.49\textwidth]{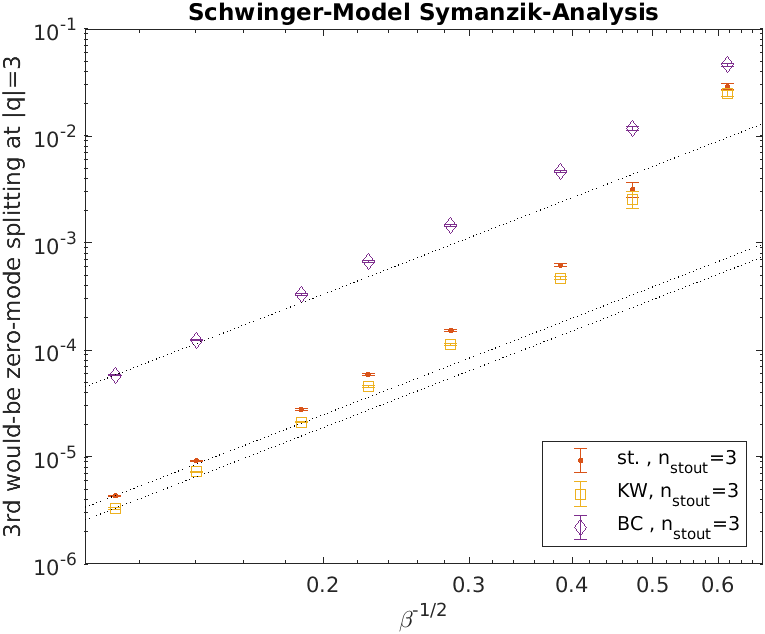}%
\caption{\label{fig:symanzik_wouldbe_n3}\sl
Same as Fig.~\ref{fig:symanzik_wouldbe_n0} but for $n_\mr{stout}=3$.
For each operator the dotted line is a power law $a\de \propto a^3$ (i.e.\ a higher power than in Fig.~\ref{fig:symanzik_wouldbe_n0}), going through the leftmost datapoint.}
\end{figure}

We shall first address the would-be zero mode splittings, followed by those of the non-topological modes.
Fig.~\ref{fig:symanzik_wouldbe_n0} displays the would-be zero mode splittings of each fermion formulation at $n_\mr{stout}=0$ as a function of $a$ in log-log representation.
For $|q|=1$ there is one splitting (top left panel), for $|q|=2$ there are two splittings (top right and middle left), and for $|q|=3$ there are three such splittings (remaining three panels).
Without smearing all formulations seem to have asymptotic behavior $a\de_j\propto a^2$ or $\de_i\propto a$ for this observable.
This is perhaps acceptable for KW and BC fermions, but it is \emph{worse than expected} for staggered fermions.

In Fig.~\ref{fig:symanzik_wouldbe_n1} the would-be zero mode splittings of each formulation at $n_\mr{stout}=1$ are plotted as a function of $a$ in log-log representation.
This time the data suggest an asymptotic scaling behavior $\de_i\propto a^2$ for \emph{all three formulations}.
There are substantial logarithmic corrections to the asymptotic behavior, even though we simulate rather fine%
\footnote{At two $\be$-values the unsmeared plaquette $1-s_\mr{wil}^{(0)}$ is above 0.99 and at three more above 0.96, see Tab.~\ref{tab:cutoff}}
Schwinger model lattices.

In Fig.~\ref{fig:symanzik_wouldbe_n3} the taste violations pertinent to would-be zero modes at $n_\mr{stout}=3$ are plotted as a function of $a$ in log-log representation.
Like in the previous figure, the would-be zero mode splittings scale as $\de_i\propto a^2$ for all three fermion operators,
but with the additional smearing steps the subleading logarithmic corrections seem even more pronounced.

Note that the dotted lines in these figures are no fits.
They show the Symanzik behavior $\de \propto a^p$ with the conjectured power $p$, starting from the leftmost data point.
This way one can gauge the size of logarithmic corrections by visual inspection.

%%%%%%%%%%

\begin{figure}[!tb]
\includegraphics[width=0.49\textwidth]{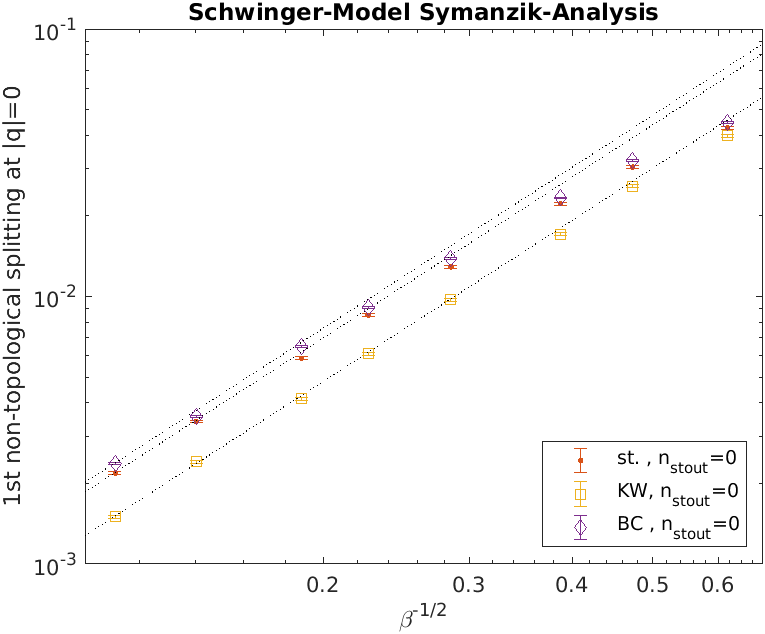}\hfill
\includegraphics[width=0.49\textwidth]{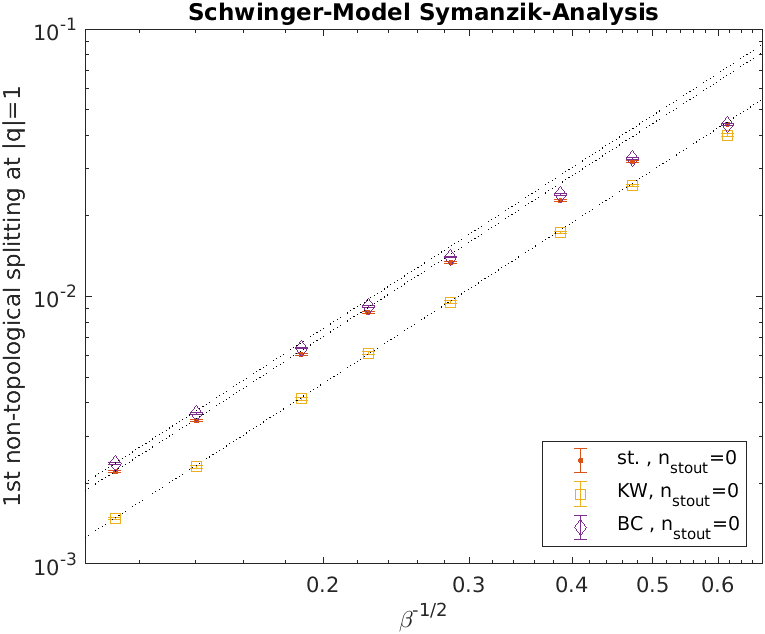}\\[2mm]
\includegraphics[width=0.49\textwidth]{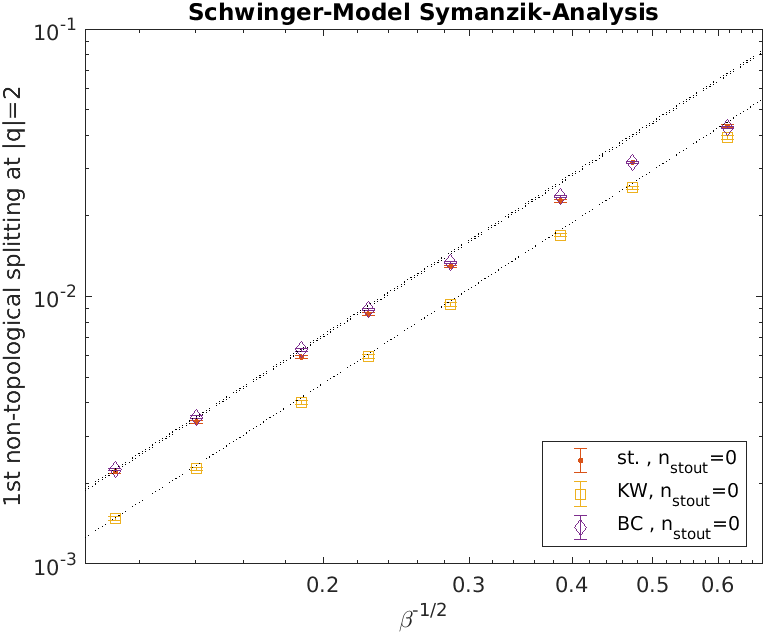}\hfill
\includegraphics[width=0.49\textwidth]{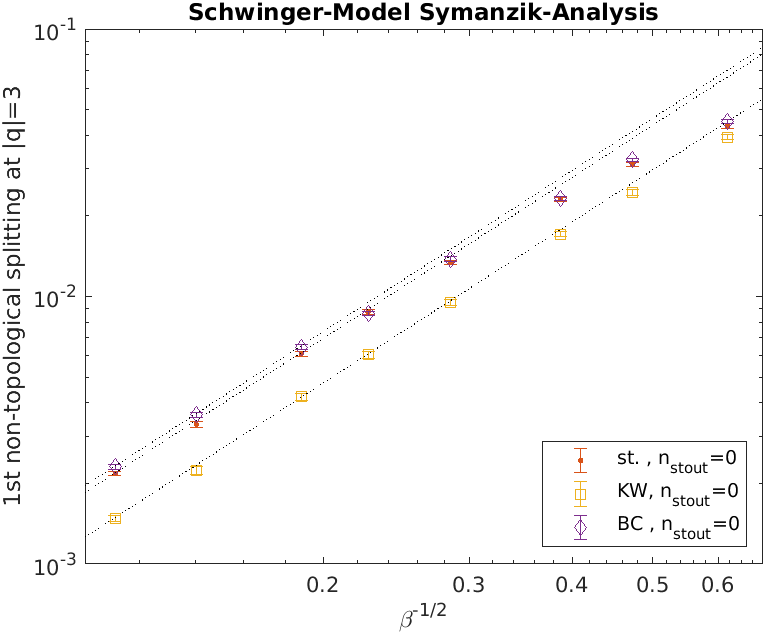}%
\caption{\label{fig:symanzik_nontop_n0}\sl
First non-topological splitting $a\de_{|q|+1}$ at $|q|=0,1,2,3$ versus $a$ for $n_\mr{stout}=0$.
Each dotted line represents a power law $a\de \propto a^2$, adjusted to go through the leftmost datapoint.}
\end{figure}

\begin{figure}[!tb]
\includegraphics[width=0.49\textwidth]{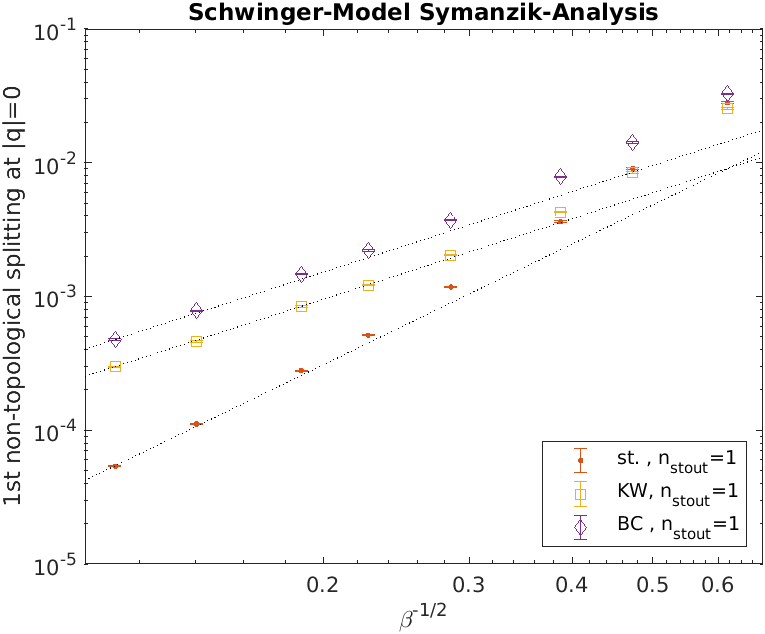}\hfill
\includegraphics[width=0.49\textwidth]{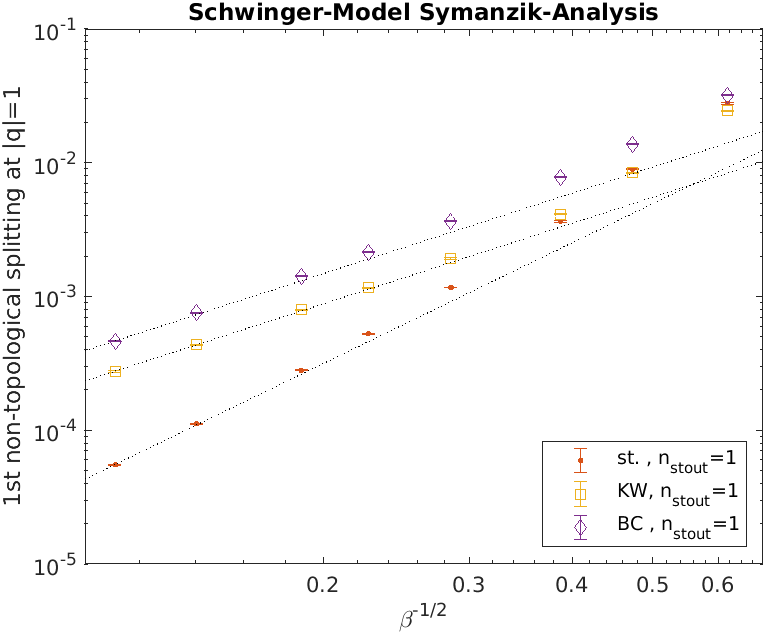}\\[2mm]
\includegraphics[width=0.49\textwidth]{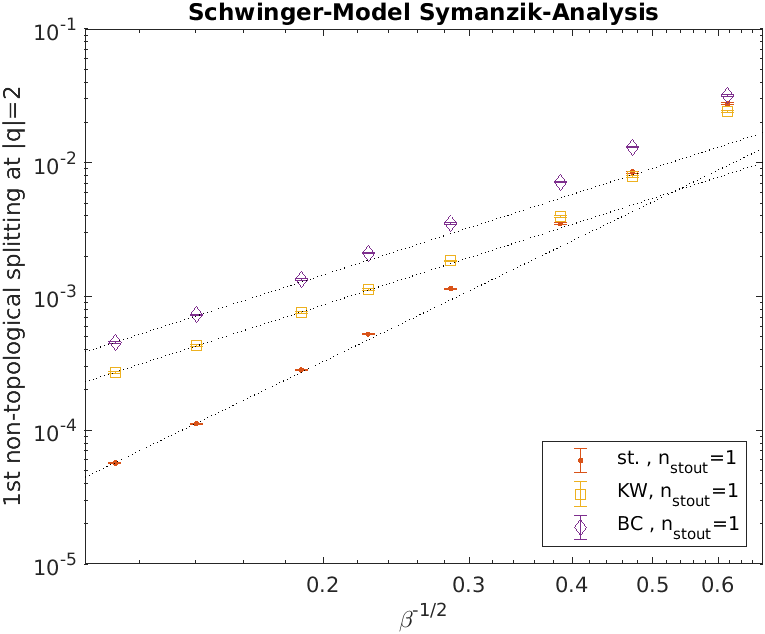}\hfill
\includegraphics[width=0.49\textwidth]{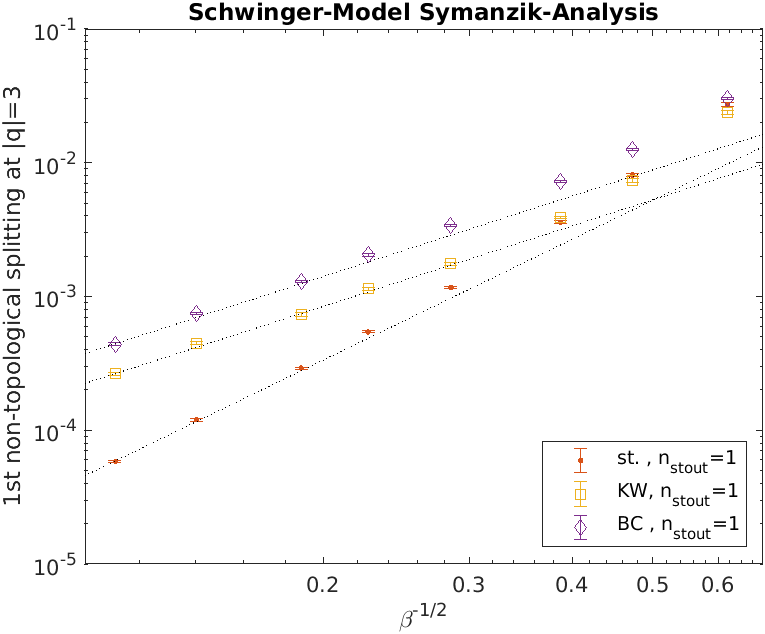}%
\caption{\label{fig:symanzik_nontop_n1}\sl
Same as Fig.~\ref{fig:symanzik_nontop_n0} but for $n_\mr{stout}=1$.
For $\DS$ the dotted line is a power law $a\de \propto a^3$, for $\DKW$ and $\DBC$ it is $a\de \propto a^2$, always adjusted to the leftmost datapoint.}
\end{figure}

\begin{figure}[!tb]
\includegraphics[width=0.49\textwidth]{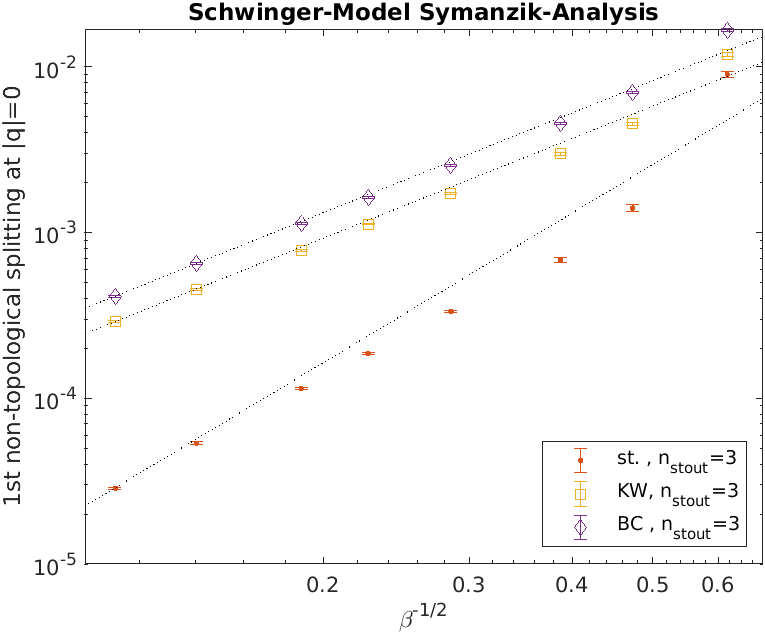}\hfill
\includegraphics[width=0.49\textwidth]{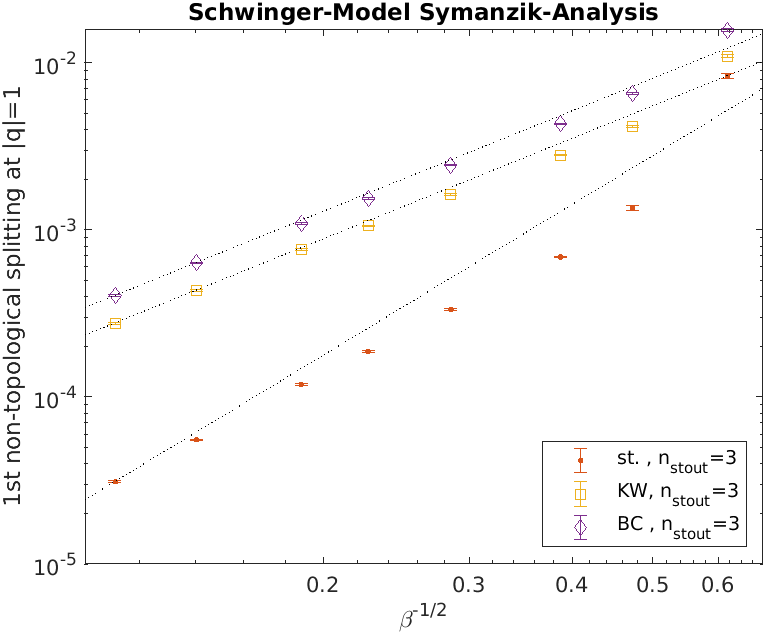}\\[2mm]
\includegraphics[width=0.49\textwidth]{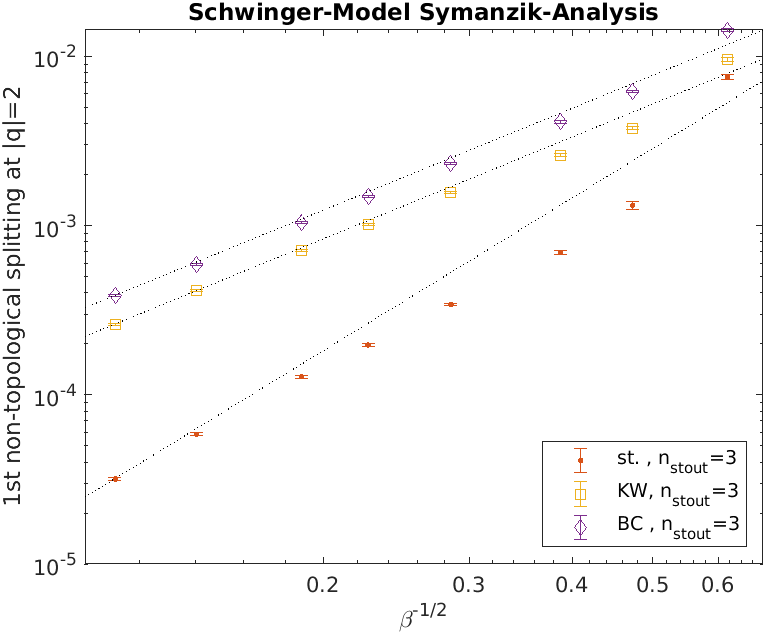}\hfill
\includegraphics[width=0.49\textwidth]{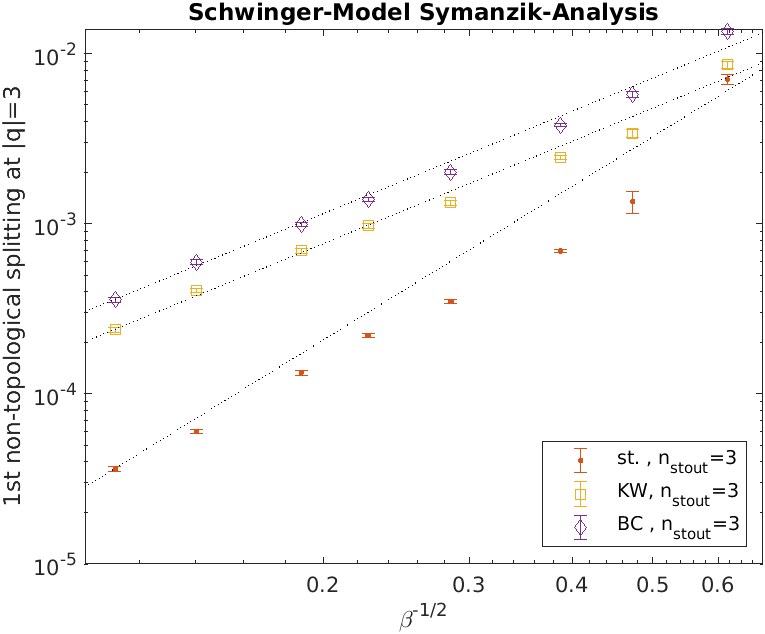}%
\caption{\label{fig:symanzik_nontop_n3}\sl
Same as Fig.~\ref{fig:symanzik_nontop_n0} but for $n_\mr{stout}=3$.
For $\DS$ the dotted line is a power law $a\de \propto a^3$, for $\DKW$ and $\DBC$ it is $a\de \propto a^2$, always adjusted to the leftmost datapoint.}
\end{figure}

We now turn to the non-topological modes.
Fig.~\ref{fig:symanzik_nontop_n0} displays these splittings for each fermion formulation at $n_\mr{stout}=0$ as a function of $a$ in log-log representation.
For $q=0$ it is $a\de_1=a\la_2-a\la_1$ (top left panel), for $|q|=1$ it is $a\de_2=a\la_3-a\la_2$ (top right panel), and analogously for $|q|=2$ (lower left panel) and $|q|=3$ (lower right panel).
Without smearing all formulations seem to have asymptotic behavior $a\de_j\propto a^2$ or $\de_i\propto a$ for this observable.
Again, this seems acceptable for KW and BC fermions, but it is \emph{worse than expected} for staggered fermions.

In Fig.~\ref{fig:symanzik_nontop_n1} results for taste violations pertinent to non-topological modes at $n_\mr{stout}=1$ are plotted as a function of $a$ in log-log representation.
Here the asymptotic behavior depends on the Dirac operator; we find $\de_i\propto a^2$ for $\DS$ and $\de_i\propto a$ for $\DKW,\DBC$.
The data suggest that there are substantial logarithmic corrections to the asymptotic behavior for $\be<20$.

In Fig.~\ref{fig:symanzik_nontop_n3} the same information is shown with $n_\mr{stout}=3$ smearings.
Results are similar to those in the previous figure, except that the sign of the logarithmic corrections seems reversed,
and the staggered non-topological splittings require truly large $\be$-values, at this level of smearing, to show asymptotic Symanzik scaling.

\begin{table}[!tb]
\centering
\begin{tabular}{|l|c@{\,}c|c@{\,}c|}
\hline
                           & $n_\mr{stout}$ & $=0$ & $n_\mr{stout}$ & $=1,3$ \\
                           & wbz            & ntm  & wbz            & ntm    \\
\hline
$\de_\mr{stag}\propto a^p$ &        1       &   1  &        2       &    2   \\
$\de_\mr{KW}  \propto a^p$ &        1       &   1  &        2       &    1   \\
$\de_\mr{BC}\;\propto a^p$ &        1       &   1  &        2       &    1   \\
\hline
\end{tabular}
\caption{\sl\label{tab:symanzik}
Power $p$ in the asymptotic Symanzik scaling law $\de \propto a^p$, as suggested by our data for staggered, KW and BC fermions.
The behavior without and with link smearing, and for would-be zero modes (``wbz'') versus non-topological modes (``ntm'') is listed separately.}
\end{table}

Clearly, the most puzzling observation is that the Symanzik power of staggered taste breakings seems to depend on the smearing level.
Without smearing we see $\de_j^\mr{st}\propto a$, with smearing we find $\de_j^\mr{st}\propto a^2$.
For any fixed $(\rh,n_\mr{stout}$) combination the smearing amounts to an \emph{ultralocal modification} of $\DS$; the asymptotic behavior should be insensitive to such a modification.
Of course, it is conceivable that any of our Figs.~\ref{fig:symanzik_wouldbe_n0}--\ref{fig:symanzik_nontop_n3} does not reflect the true asymptotic behavior,
but in view of the large $\be$-values this would be surprising.
For further discussion we summarize the conjectured powers $p$, as suggested by our data, in Tab.~\ref{tab:symanzik}.

%%%%%%%%%%%%%%%%%%%%%%%%%%%%%%%%%%%%%%%%%%%%%%%%%%%%%%%%%%%%%%%%%%%%%%%%%%%%%%%%

\section{Conclusions and outlook \label{sec:con}}

%%%%%%%%%%%%%%%%%%%%%%%%%%%%%%%%%%%%%%%%%%%%%%%%%%%%%%%%%%%%%%%%%%%%%%%%%%%%%%%%

The goal of this paper has been to assess how a continuum Dirac operator eigenvalue is split into a pair of near-degenerate eigenvalues
for staggered, Karsten-Wilczek (KW) and Bori\c{c}i-Creutz (BC) fermions in 2D.
On typical gauge backgrounds staggered taste splittings were found to diminish \emph{exponentially} in the gradient flow time $\ta$,
provided the latter is not too large (see Fig.~\ref{fig:gradflow_deltas} and Ref.~\cite{Ammer:2022ksl}).
At a given level of smearing (or gradient flow time in lattice units; we chose $\ta/a^2=0,0.25,0.75$) the taste-breaking effects were found to disappear in the continuum limit,
albeit not necessarily as fast as standard arguments would suggest.

We find that it makes a difference whether the underlying continuum mode is a would-be zero mode (``wbz'') or a non-topological (``ntm'') mode.
For wbz-modes the splitting disappears as $\de_\mr{wbz} \propto a$ for all three actions without link smearing, and $\de_\mr{wbz} \propto a^2$ for all three actions with smearing.
This observation is disturbing, as it challenges the standard view \cite{Blum:1996uf,Orginos:1998ue,Lagae:1998pe,Lepage:1998vj,Knechtli:2000ku} that a fixed level of smearing does not change the Symanzik universality class of a given fermion operator.

For non-topological modes we end up with the unexpected observation that only the staggered action with link smearing scales asymptotically as $\de_\mr{ntm} \propto a^2$,
while our data for unsmeared staggered fermions and for KW and BC with and without link smearing suggest $\de_\mr{ntm} \propto a$.
The conjectured powers in the asymptotic Symanzik law $\de \propto a^p$ are summarized in Tab.~\ref{tab:symanzik},
and we recall that logarithmic corrections were seen in all cases, in particular with smearing.

We emphasize that our results for KW and BC fermions are for the bare actions, i.e.\ without marginal counterterms (see e.g.\ Ref.~\cite{Weber:2023kth} for details).
Including such counterterms might change some of the entries in the lower two lines of Tab.~\ref{tab:symanzik}.
Unfortunately, the coefficients of these counterterms are not known in 2D (neither perturbatively nor non-perturbatively, neither with nor without stouting).
For staggered fermions there are no such counterterms, and the conjectured change from $p=1$ to $p=2$ with smearing remains mysterious.

Evidently, our findings are unexpected and deserve due diligence.
It is conceivable that our data do not reflect the true asymptotic behavior.
This, however, would be surprising, since our lattices are rather fine compared to previous investigations
(we bridge a factor $\be_\mr{max}/\beta_\mr{min}=25$, tantamount to a factor $5$ between the largest/smallest lattice spacing).
Assuming there is no technical issue with our code%
\footnote{Of course, we started scrutinizing our code.
We mentioned in Sec.~\ref{sec:sim} that we checked our simulation data against analytic results of Ref.~\cite{Elser:2001pe}.
We carefully checked the implementation of the fermion operators against those used in Ref.~\cite{Durr:2022mnz}.
Also the smearing seems fine; when applied to a purely gluonic observable it seems to imply beautiful Symanzik scaling at each smearing level (including $n_\mr{stout}=0$), see App.~\ref{sec:app} for details.
Last but not least, the eigenvalue determination could be flawed.
But we use canned routines, and we checked that the functions {\tt eig} and {\tt eigs} in Matlab (which use different algorithms) yield identical results.}
it seems that the interplay between smearing and subleading logarithmic corrections to asymptotic Symanzik scaling deserves a closer look (based on precise data).
For a discussion of logarithmic corrections see Refs.~\cite{Balog:2009yj,Balog:2009np,Balog:2012db,Husung:2019ytz,Husung:2021mfl,Husung:2022kvi}.

Several research strategies may shed some light on the conundrum presented in this paper.
First, it would be interesting to know whether the $|q|$ real-valued eigenvalues in the physical branch of an unimproved Wilson fermion (which belong to ``would-be zero modes'') vanish in the continuum limit like $\la_\mr{wbz}\propto a$,
and whether tree-level or one-loop improvement would change this behavior to $\la_\mr{wbz}\propto a\log(a)$ or $\la_\mr{wbz}\propto a\log^2(a)$, as the Symanzik surmise suggests.
Again, one would not expect to find any dependence on the amount of smearing (keeping the flow time fixed in lattice units, $\ta/a^2=\mr{const}$), but surprises may happen.
Second, it would be interesting to repeat the analysis of this paper with a spectroscopy based measure of the taste splitting (defined as the difference
between the squared pion masses with different taste structure), for all three fermion formulations, and to check whether the results depend on the amount of link smearing.
With smearing one should be prepared to see large logarithmic corrections, to the point that it becomes a challenge to determine the asymptotic Symanzik power $p$.

%%%%%%%%%%%%%%%%%%%%%%%%%%%%%%%%%%%%%%%%%%%%%%%%%%%%%%%%%%%%%%%%%%%%%%%%%%%%%%%%

\appendix

%%%%%%%%%%%%%%%%%%%%%%%%%%%%%%%%%%%%%%%%%%%%%%%%%%%%%%%%%%%%%%%%%%%%%%%%%%%%%%%%

\section{Symanzik scaling of topological charge Z-factor \label{sec:app}}

%%%%%%%%%%%%%%%%%%%%%%%%%%%%%%%%%%%%%%%%%%%%%%%%%%%%%%%%%%%%%%%%%%%%%%%%%%%%%%%%

\begin{figure}[!tb]
\includegraphics[width=0.49\textwidth]{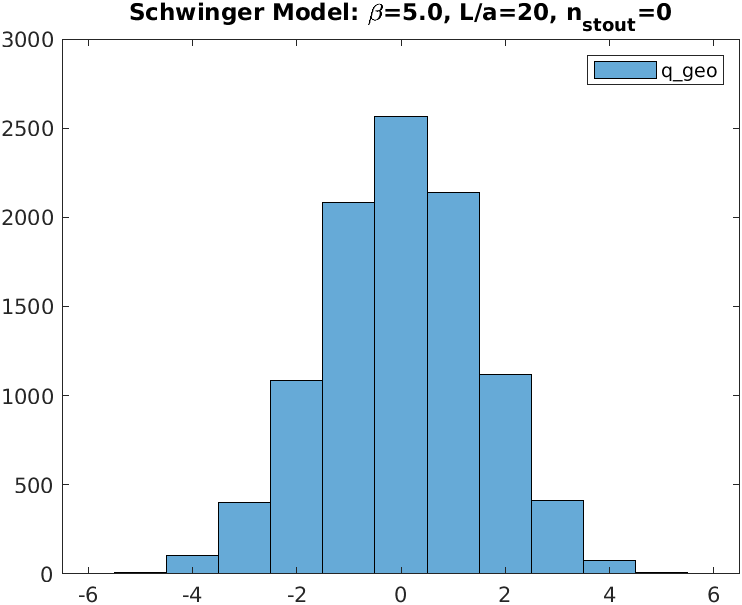}\hfill
\includegraphics[width=0.49\textwidth]{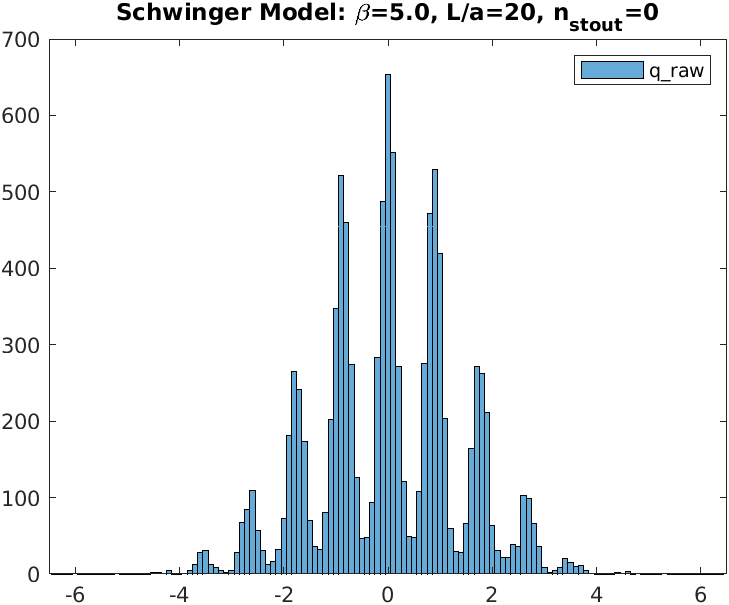}%
\vspace*{-2mm}
\caption{\label{fig:appone}\sl
Histogram of the integer valued topological charge $q_\mr{geo}$ (left), and of the real valued charge $q_\mr{raw}$ (right)
in the ensemble $\be=5.0$, $L/a=20$, $n_\mr{stout}=0$ with 10\,000 configs.}
\vspace*{+2mm}
\includegraphics[width=0.49\textwidth]{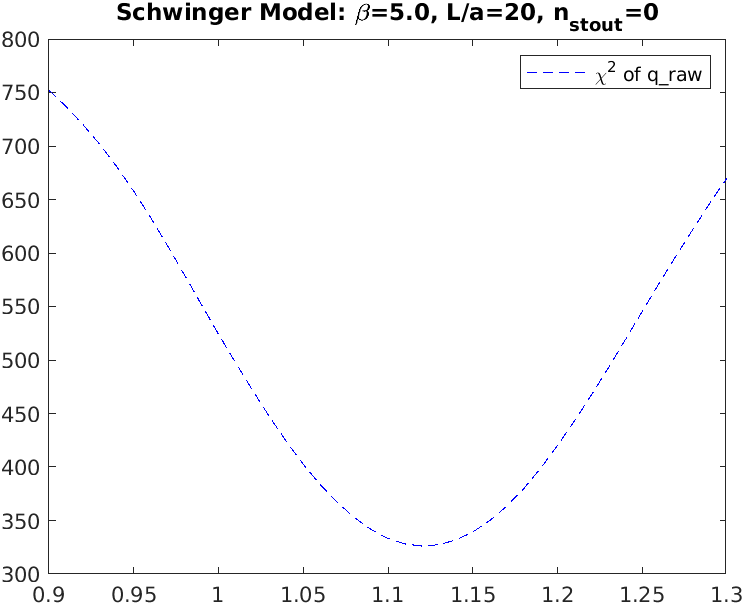}\hfill
\includegraphics[width=0.49\textwidth]{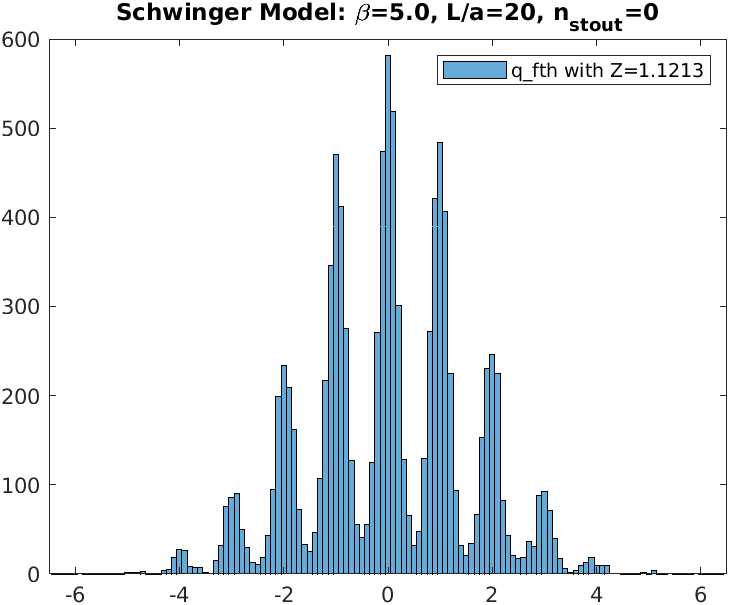}%
\vspace*{-2mm}
\caption{\label{fig:apptwo}\sl
Dependence of $\ch^2$ on $Z$ for $n_\mr{stout}=0$; the abscissa of its minimum defines $Z$ (left).
The resulting histogram of $Zq_\mr{raw}$ (right), before rounding, for the same ensemble as in Fig.~\ref{fig:appone}.}
\end{figure}

A lattice regularized topological charge density $q(x)$ or (global) charge $q$ renormalizes multiplicatively relative to its continuum counterpart \cite{Campostrini:1988cy,Campostrini:1989dh}.
The respective $Z$-factor can be read off from the histogram of $q_\mr{raw}$ as defined in (\ref{def_qraw}).
This is exemplified in Fig.~\ref{fig:appone} for $(\be,n)=(5.0,0)$.
For instance, if the peaks are near $\pm0.9,\pm1.8,\pm2.7,...$, then the choice $Z=1/0.9\simeq1.11$ would make $Zq_\mr{raw}$ peak near integer-valued numbers.
To formalize this observation, one defines
\beq
\chi^2(Z)=\sum_U \Big(\mr{round}(Zq_\mr{raw}^{(n)}[U])-Zq_\mr{raw}^{(n)}[U] \Big)^2
\eeq
where the sum is over all configurations in a given ensemble.
The local minimum of $\ch^2$ at $Z>1$ defines $Z$, and this renormalization condition was also used in Ref.~\cite{DiGiacomo:1991ba,Alles:1993ij,DelDebbio:2002xa,Durr:2006ky}.
For $(\be,n)=(5.0,0)$ Fig.\,\ref{fig:apptwo} indicates that $Z\simeq1.12$ indeed aligns the peaks in the histogram to integer values.

\begin{table}\centering
\begin{tabular}{|l|c@{\,\,}c@{\,\,}c@{\,\,}c@{\,\,}c@{\,\,}c@{\,\,}c@{\,\,}c|}
\hline
$\beta$     &    3.2   &    5.0   &   7.2   &   12.8  &   20.0  &   28.8  &   51.2  &   80.0  \\
$L/a$       &     16   &    20    &    24   &    32   &    40   &    48   &    64   &    80   \\
\hline
%%%
%%% grep "Z_opt="      splittings_2D_ana.diary_stouting | awk '{print $2}'
%%% remove finite-volume part
%%% obj=[...]; obj=reshape(obj,3,8)
%%%       1.2032       1.1213       1.0776       1.0412       1.0262       1.0177       1.0099       1.0064
%%%       1.0587       1.0273       1.0170       1.0094       1.0061       1.0042       1.0024       1.0016
%%%       1.0183       1.0089       1.0060       1.0035       1.0023       1.0016       1.0009       1.0006
%%% grep "Z_opt: err=" splittings_2D_ana.diary_stouting | awk '{print $3}'
%%% remove finite-volume part
%%% obj=[...]; obj=reshape(obj,3,8)
%%%    0.0050690   0.00130260   0.00077692   0.00036338   0.00022696   0.00015248   8.3805e-05   5.2498e-05
%%%    0.0014767   0.00044277   0.00011034   4.6138e-05   3.1057e-05    2.088e-05   1.2071e-05   7.7422e-06
%%%    0.0010842   0.00018395   5.0384e-05   1.6395e-05   1.1411e-05   7.6709e-06   4.4395e-06   2.9805e-06
%%%
$Z^{(n=0)}$ &1.2032(50)&1.1213(13)&1.0776(7)&1.0412(4)&1.0262(2)&1.0177(2)&1.0099(1)&1.0064(1)\\
$Z^{(n=1)}$ &1.0587(15)&1.0273(04)&1.0170(1)&1.0094(0)&1.0061(0)&1.0042(0)&1.0024(0)&1.0016(0)\\
$Z^{(n=3)}$ &1.0183(11)&1.0089(02)&1.0060(1)&1.0035(0)&1.0023(0)&1.0016(0)&1.0009(0)&1.0006(0)\\
\hline
\end{tabular}
\caption{\label{tab:z_cutoff}\sl
Results for $Z^{(n)}$ in the ``cut-off effect'' study as defined in Tab.~\ref{tab:cutoff}.
Every column comprises three ensembles of 10\,000 configurations each, subject to either 0, 1 or 3 stout steps.}
\end{table}

\begin{table}\centering
\begin{tabular}{|l|ccccc|}
\hline
$\beta$     &   7.2   &   7.2   &   7.2   &   7.2   &   7.2   \\
$L/a$       &    16   &    20   &    24   &    32   &    40   \\
\hline
$Z^{(n=1)}$ &1.0171(2)&1.0170(1)&1.0170(1)&1.0171(1)&1.0172(1)\\
\hline
\end{tabular}
\caption{\label{tab:z_finvol}\sl
Results for $Z^{(1)}$ in the ``finite volume'' study as defined in Tab.~\ref{tab:finvol}.
Each measurement uses 10\,000 configurations and a single stout smearing step.}
\end{table}

In Tabs.~\ref{tab:z_cutoff} and \ref{tab:z_finvol} we summarize our results for the topological charge renormalization factor $Z$ for all of our ensembles.
The results depend on $\be$ and the smearing level $n$; increasing either one drives the value closer to $1$.
After Symanzik \cite{Symanzik:1983dc,Symanzik:1983gh,Curci:1983an,Luscher:1984xn} one expects that
\beq
Z(\be,n)=1+\mr{const}_n\,a^2+O(a^4)
\label{z_symanzik}
\eeq
since both the gauge action (\ref{def_swil}) and the operator (\ref{def_qraw}) admit leading cut-off effects $\propto a^2$.
Our choice of setting the lattice spacing $a$ through the gauge coupling $e$ implies that one may replace $a^2$ by $1/\be$ in the Schwinger model.
The results tabulated suggest that $Z$ is indeed independent of the volume, whereupon the hypothesis (\ref{z_symanzik}) contains all relevant dependencies.

\begin{figure}[!tb]
\includegraphics[width=0.49\textwidth]{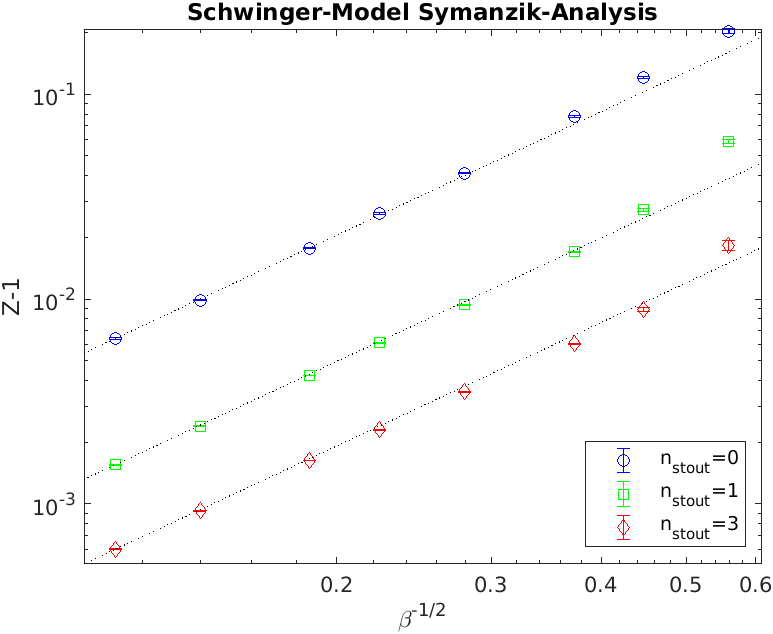}\hfill
\includegraphics[width=0.49\textwidth]{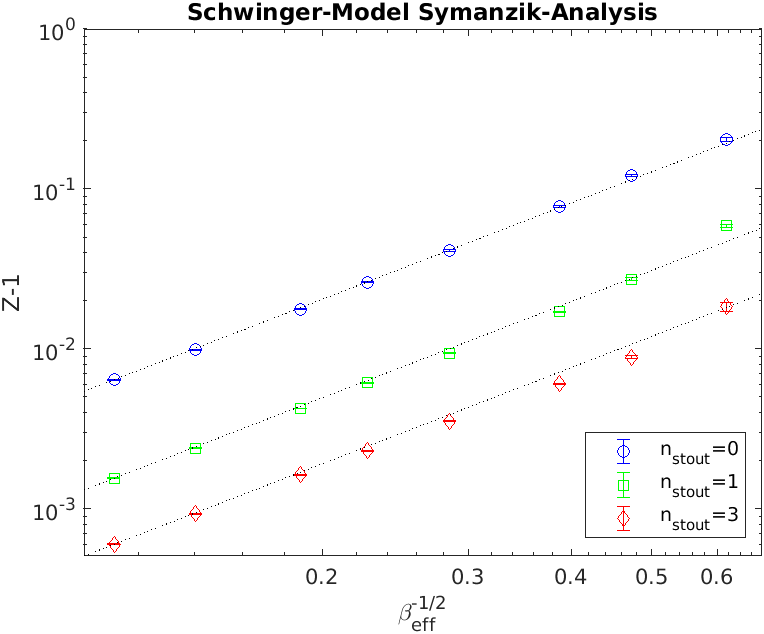}%
\caption{\label{fig:symanzik_z}\sl
$Z^{(n)}-1$ versus two varieties of $a$ (left $\be^{-1/2}$, right $\be_\mr{eff}^{-1/2}$, see text), for $n_\mr{stout}=0,1,3$.
The dotted lines are power laws $\propto a^2$ passing through the leftmost datapoint.}
\end{figure}

In Fig.~\ref{fig:symanzik_z} the quantity $Z-1$ is plotted as a function of $\be^{-1/2}$ in log-log representation.
The data support the Symanzik scaling hypothesis (\ref{z_symanzik}), with a prefactor which depends on $n$.
The dotted lines are no fits; they implement the Symanzik power $a^2$, with a prefactor adjusted to make them (exactly) pass through the leftmost (most continuum-like) datapoint.
The left panel shows that it takes large $\be$-values to see good agreement with Symanzik scaling.
Ref.~\cite{deForcrand:1997fm} proposes to use $\be_\mr{eff}=\<U_\dal\>\be$ to set the lattice spacing in the Schwinger model (this shifts all datapoints a bit to the right).
The right panel shows that this improves things slightly for $n_\mr{stout}=0$, but the asymptotic behavior (to the far left) is unchanged.

Overall, both panels illustrate the applicability of the Symanzik scaling hypothesis (\ref{z_symanzik}) for the quantity $Z-1$ at each smearing level $n_\mr{stout}\in\{0,1,3\}$.
The reluctance to assume Symanzik scaling and the dependence of the power $p$ on $n_\mr{stout}$ that were reported in the main investigation are thus genuine to the fermion
operators, and unrelated to the smearing procedure.

%%%%%%%%%%%%%%%%%%%%%%%%%%%%%%%%%%%%%%%%%%%%%%%%%%%%%%%%%%%%%%%%%%%%%%%%%%%%%%%%

%%%%%%%%%%%%%%%%%%%%%%%%%%%%%%%%%%%%%%%%%%%%%%%%%%%%%%%%%%%%%%%%%%%%%%%%%%%%%%%%

\end{document}